\documentclass[article]{IEEEtran}
\usepackage{cite}

\usepackage{amsmath,amssymb,amsfonts}
\usepackage{mathtools}

\usepackage{algorithmic}
\usepackage{graphicx}
\usepackage{textcomp}
\usepackage{xcolor}
\usepackage{hyperref}
\usepackage{mathrsfs}
\usepackage{amsthm}
\usepackage{scalerel,stackengine}
\usepackage{paralist}
\usepackage{blkarray}
\usepackage[ruled,vlined,linesnumbered]{algorithm2e}
\usepackage{empheq}
\definecolor{lightgreen}{HTML}{90EE90}

\usepackage{amsmath}
\usepackage[most]{tcolorbox}

\tcbset{colback=blue!3!white, colframe=red!50!black,
        highlight math style= {enhanced, 
     colframe=red,colback=red!10!white,boxsep=0pt}
       }

    \usepackage[compact]{titlesec}
    \titlespacing{\section}{0pt}{2.5ex}{1ex}
    \titlespacing{\subsection}{0pt}{1ex}{0ex}

\usepackage{newtxmath}
\usepackage{alltt}
\usepackage{epsfig}
\usepackage{verbatim}
\usepackage{fancybox}
\usepackage{subfigure}
\usepackage{url}
\usepackage{multirow}
\usepackage{setspace}
\usepackage{fancyhdr}
\usepackage{framed}
\usepackage{booktabs}
\usepackage{color}
\usepackage{wrapfig}
\usepackage{mathtools}
\mathtoolsset{showonlyrefs}
\usepackage{placeins}

\usepackage{tikz}
\usepackage{enumitem}
\setitemize{noitemsep,topsep=0pt,parsep=0pt,partopsep=0pt}

\newcommand\oast{\stackMath\mathbin{\stackinset{c}{0ex}{c}{0ex}{\ast}{\bigcirc}}}
\def\delequal{\mathrel{\ensurestackMath{\stackon[1pt]{=}{\scriptstyle\Delta}}}}

\allowdisplaybreaks

\newtheorem{remark}{Remark}

\newtheorem{definition}{Definition}

\newtheorem{assumption}{Assumption}
\newtheorem{result}{Result}

\begin{document}

	\title{The Companion Model\textemdash a Canonical Model in Graph Signal Processing}
\author{John Shi, \IEEEmembership{Student Member, IEEE,} Jos\'e M.~F.~Moura, \IEEEmembership{Fellow, IEEE}%
\thanks{This material is based upon work partially funded and supported by the Department of Defense under Contract No. FA8702-15-D-0002 with Carnegie Mellon University for the operation of the Software Engineering Institute, a federally funded research and development center. This work is also partially supported by NSF grants CCF~1837607 and CCN~1513936.}%
\thanks{Department of Electrical and Computer Engineering, Carnegie Mellon University, Pittsburgh PA 15217 USA; [jshi3,moura]@andrew.cmu.edu.}}

	\maketitle
	\begin{abstract}
%
%
This paper introduces a \textit{canonical} graph signal model defined by a \textit{canonical} graph and a \textit{canonical} shift, the \textit{companion} graph and the \textit{companion} shift. These are canonical because, under standard conditions, we show that any graph signal processing (GSP) model can be transformed into the canonical model. The transform that obtains this is the graph $z$-transform ($\textrm{G$z$T}$) that we introduce. The GSP canonical model comes closest to the discrete signal processing (DSP) time signal models: the structure of the companion shift decomposes into a line shift and a signal continuation just like the DSP shift and the GSP canonical graph is a directed line graph with a terminal condition reflecting the signal continuation condition. We further show that, surprisingly, in the canonical model, convolution of graph signals is fast convolution by the DSP FFT.
\end{abstract}
\vspace{-.3cm}
\section{Introduction}\label{sec:intro}
In DSP, a sample $s_n$ of a real valued time signal like a sample of a segment of speech or of audio is indexed by the time instant~$n$ at which the sample occurs. Similarly, the intensity or color $s_{ij}$ of an image pixel is indexed by the location $(i,j)$ of the pixel. Both indices, time~$n$ and pixel $(i,j)$, usually take values on regularly spaced one dimensional (1D) or two dimensional (2D) grids.\footnote{We will assume in this paper that the index sets are finite.} In contrast, the samples of graph signals like temperature readings in a network of meteorological stations or like voltages in a electric grid are indexed by the weather stations or by the buses of the power grid network, seemingly placed in arbitrary locations in space. This indexing structure is better described by a graph. Graph Signal Processing (GSP) \cite{Sandryhaila:13,Sandryhaila:14,Sandryhaila:14big,ShumanNFOV:13} develops methods to analyze and study data indexed or defined on graphs, extending Discrete Signal Processing (DSP) to these signals \cite{siebert-1986,oppenheimwillsky-1983,oppenheimschaffer-1989,mitra-1998}.

Like with DSP, GSP commonly describes the graph signals by their values $s_n$ at the~$N$  vertices~$n=0,\cdots,N-1$ of the indexing graph~$G=(V,E)$, with vertex set~$V$ and edge set~$E$. In GSP, the adjacency matrix~$A$ of~$G$ is taken as the shift \cite{Sandryhaila:13,Sandryhaila:14,Sandryhaila:14big}. The shift~$A$ is the basic building block of linear shift invariant (LSI) filters that are polynomials $P(A)$ of~$A$. Alternatively, the signals can be described by their graph spectrum $\widehat{s}=\textrm{GFT} \,s$, where $\textrm{GFT}$ is the graph Fourier transform. The $\textrm{GFT}$ is the inverse of the matrix of eigenvectors of~$A$ and diagonalizes the shift. In \cite{shimoura-2021}, when studying graph sampling, we introduced a spectral graph shift~$M$. This led us to consider a new ``spectral'' graph signal model where the spectral LSI filters are polynomials $P(M)$ in~$M$ rather than in~$A$. The shift~$M$ \cite{shimoura-2021} replicates for GSP, with appropriate interpretation, the DSP property that shifting a signal in one domain, phase shifts the signal in the other domain, and vice-versa.

The vertex signal~$s$ and the graph spectrum~$\widehat{s}$, are two ways of describing the same graph signal but with respect to two different bases, the vertex standard Euclidean basis and the graph Fourier basis. Each of these signal descriptions has its own advantages. The vertex basis is the natural one, since the data is often collected at each node. The graph Fourier basis decomposes the signal model space into invariant subspaces (for diagonalizable shifts, these are the~$N$ one-dimensional eigenvector spaces) and signals aligned with these invariant subspaces are invariant to linear graph filtering.

\textbf{Contributions}. This paper considers the following question: Is there, for arbitrary indexing graph~$G$, a signal model that replicates as closely as possible the DSP signal model. We interpret this question in terms of the action of the shift~$A$ on the signal~$s$. In DSP, $A\,s$ is a delayed signal, i.e., the signal samples are shifted downwards and then the signal is wrapped around so that the last signal sample $s_{N-1}$ reappears at the top of $A\,s$. This is known as periodic signal extension. The answer to this question for generic graph~$G$ turns out, surprisingly at first sight, to be yes. In other words, there is a graph signal representation that reproduces many of the characteristics of DSP starting with a GSP shift that moves downwards the graph signal (details in section~\ref{sec:GSPcompanionmodel}).  For reasons that will become clear in sections~\ref{sec:polynomialrep-a} and~\ref{sec:GSPcompanionmodel}, we call the signal transform that achieves this the graph $z$-transform ($\textrm{G$z$T}$), and we refer to the signal model as the graph \textit{companion} model. Surprising facts associated with this signal model include:
\begin{inparaenum}[1)]
\item the graph signal shift delays the graph signal;
   \item like in DSP, the model is defined by the graph frequencies (eigenvalues of~$A$), with no role played by the eigenvectors of~$A$;
     \item it conduces to a canonical (weighted) shift and a canonical (weighted) graph, regardless of the underlying graph~$G$; and
         \item it leads to a \textit{fast} convolution of graph signals using the DSP FFT.
             \end{inparaenum}


%
%
%
\textbf{Brief review of the literature}. The GSP literature is vast, by now covering many topics in processing graph signals. The approach in \cite{Sandryhaila:13,Sandryhaila:14,Sandryhaila:14big} identifies as basic building block the shift filter~$A$, building on the Algebraic Signal Processing in \cite{Pueschel:03a,Pueschel:05e,Pueschel:08a,Pueschel:08b,Pueschel:08c}. The approach in \cite{ShumanNFOV:13} departs from a variational operator, the graph Laplacian~$L$, motivated by for example earlier work from spectral graph theory \cite{Chung:96,Belkin:02,Coifman:05a}, from work extending wavelets to data from irregularly placed sensors in sensor networks \cite{guestrinbodikthibauxpaskinmadden-ipsn2004,wagnerbaraniuketal-SSPWorkshop2005,Hammond:11,Narang:12}, and from research on sampling graph based data \cite{Narang:10,Narang:11}. A comprehensive review covering both approaches and illustrating many different applications of GSP is \cite{ortegafrossardkovacevicmouravandergheynst-2018}.

Many additional topics have been considered in GSP. A sample of these include: alternative (unitary, but not local) shift operators \cite{giraultgoncalvesfleury-2015,gavilizhang-2017}; approximating graph signals \cite{Zhu:12}; extensive work on sampling of graph signals, e.g.,  \cite{anis2014towards,Jelena,marques2015sampling,anis2016efficient,chamon2017greedy,tanaka-2018}, see the recent review \cite{EldarTanakaSPM}; extending classical multirate signal processing to graphs \cite{tekevaidyanathan-2017-I,tekevaidyanathan-2017-II}; an uncertainty principle for graph signals \cite{Agaskar:12}; the study of graph diffusions \cite{pasdeloupgriponmercierpastorrabbat-2018}; graph signal recovery \cite{chen2015signal-2}; interpolation and reconstruction of graph signals \cite{segarra2015interpolation,segarra2016reconstruction}; stationarity of graph processes \cite{marquessegarraleusribeiro-17}; learning graphs from data \cite{meimoura-2017,meimoura-2018,dongthanourabbatfrossard-2019}; or non-diagonalizable shifts and the graph Fourier transform \cite{derimoura-2017}.

All the references above describe graph signals by their standard (node or vertex) representation~$s$ or their spectral representation~$\widehat{s}$, but practically none has discussed or studied other graph signal representations or the issues related to signal representations that we pursue here.

\textbf{Summary of the paper}. Section~\ref{sec:background} casts DSP in the graph framework, provides background on GSP, and introduces graph impulses both in the vertex and the spectral domains. Section~\ref{sec:vertexgraphsignalrep} introduces signal representations in general and then the two common ones, the vertex, standard, or Euclidean signal representation and the Fourier or spectrum representation. Section~\ref{sec:polynomialrep-a} introduces the vertex impulsive representation and the graph $z$-transform. Section~\ref{sec:GSPcompanionmodel} shows that the vertex impulsive representation leads to a canonical shift, the \textit{companion} shift, and a canonical directed graph, the \textit{companion} graph. These replicate the structure of the cyclic shift and the cyclic (time) graph, with appropriate boundary condition, given by the Cayley-Hamilton Theorem. The graph \textit{companion} model comes closest to many DSP concepts. In particular, it requires only knowledge of the eigenfrequencies of~$A$, not its eigenvectors. Section~\ref{sec:repwidehats} extends the graph \textit{companion} model for~$s$ to its spectrum~$\widehat{s}$. Section~\ref{sec:signalrepdomains} summarizes the relations between the different GSP signal domains and shows how these GSP models coalesce into only two for DSP. Section~\ref{sec:graphconvolution} uses the companion signal model to introduce a fast convolution for graph signals using the DSP FFT. Finally, section~\ref{sec:conclusion} presents concluding remarks.
\section{GSP Background}
\label{sec:background}
\label{sec:notation}
This section reviews briefly GSP following \cite{Sandryhaila:13,Sandryhaila:14,Sandryhaila:14big}. Let $G=(V,E)$ be a graph of order~$N$, i.e., with vertex or node set~$V$ of cardinality $|V|=N$, and with edge set~$E$. The graph~$G$ is arbitrary, possibly  directed, undirected, or mixed with directed and undirected edges. The graph can be specified by an adjacency matrix~$A$, where $A_{ij}=1$ if there is a directed edge from node~$j$ to node~$i$ or $A_{ij}=0$ otherwise.\footnote{\label{ftn:Aij-computerscience} Computer Science reverses this convention and the adjacency is $A^T$.}
\begin{remark}[$A$~equivalence class] \label{rmk:Aequivalenceclass} The adjacency matrix~$A$ depends on the ordering of the nodes of~$V$. Different node orderings are related by permutations~$P$ and the corresponding adjacency matrices are conjugated by~$P$, i.e., $P^{-1}AP=P^TAP$. In other words, adjacency matrices describing the same graph are an equivalence class under the symmetric group of permutations. We assume that a representative of this class has been chosen, by fixing the labeling order of the nodes in~$V$, which becomes now an ordered set, and identify graph~$G$ with adjacency matrix~$A$ rather than with the class of adjacency matrices.\footnote{\label{ftn:isomorphocgraphs} Graphs that differ by permutations~$P$ of~$V$ are isomorphic.}
\end{remark}

A graph signal~$s$ is an (ordered) $N$-tuple $s=\left(s_0,\cdots,s_{N-1}\right)$ that assigns to each node $n\in V$, $n=0,\cdots,N-1$, the graph sample $s_n$. In other words, graph signal samples $s_n$ are indexed by the nodes~$n$ of the graph. In this paper, we consider the graph samples to be complex valued,  $s_n\in \mathbb{C}$. The graph signal~$s$ is then a vector in $\mathbb{C}^N$, the $N$-dimensional vector space over the complex field~$\mathbb{C}$.

\subsection{DSP as GSP}\label{subsec:dspasgsp}
\begin{figure}[htb!]
\begin{center}
	\includegraphics[width=8.5cm, keepaspectratio]{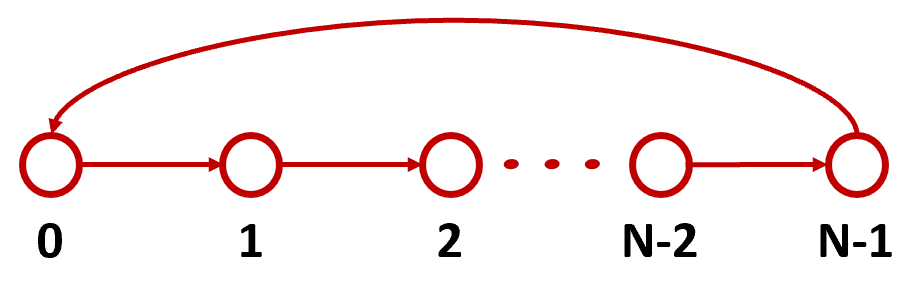}
	\caption{Directed cycle graph.}
		\label{fig:directedcyclegraph}
\end{center}
\end{figure}

 To motivate GSP, we start by casting DSP in the context of GSP, see for example \cite{Sandryhaila:13,Sandryhaila:14big}. Consider the~$N$ node directed cycle graph~$G$ in figure~\ref{fig:directedcyclegraph} with adjacency matrix~$A_c$
\begin{align} \label{eqn:Adsp}
    A_c {}&= \begin{bmatrix}
	0  & 0  & \hdots & 0 & 1 \\
	1  & 0  &  \hdots & 0 & 0 \\
	\vdots  & 1 & \ddots & \vdots  & \vdots\\
	\vdots  & \vdots & \ddots & \ddots  & \vdots\\
	0  & 0 &  \hdots & 1 & 0 \\
    \end{bmatrix}.
\end{align}
The nodes of the cycle graph~$G$ represent the time ticks~$n$ and are naturally ordered. The time signal samples~$s_n$ are indexed by the nodes of~$G$. Matrix~$A_c$ is also the matrix representation of the shift $z^{-1}$ in DSP (assuming periodic boundary conditions \cite{Pueschel:05e,Pueschel:08a,Sandryhaila:13})
\begin{align}\label{eqn:periodicshift-1}
    A_c\left[\begin{array}{c}
    s_0\\
    s_1\\
    \cdots\\
    s_{N-1}\end{array}\right]{}&=\left[\begin{array}{c}
    s_{N-1}\\
    s_0\\
    \cdots\\
    s_{N-2}\end{array}\right].
\end{align}
%
The eigenvalues (or a normalized version) $\lambda_k$  and the eigenvectors $v_k$ of the cyclic $A_c$ in~\eqref{eqn:Adsp} are the discrete time frequencies and the discrete time harmonics, spectral components, or eigenmodes of time signals
\begin{align}
\label{eqn:dspLambda-1}
\lambda_k{}&=e^{-j\frac{2\pi}{N}k},\:\:, k=0,\cdots, N-1\\
\label{eqn:dspLambda-2}
v_k{}&=\frac{1}{\sqrt{N}}\left[1 \,\,e^{j\frac{2\pi}{N}k}\,\cdots\,\, e^{j\frac{2\pi}{N}k(N-1)}\right]^T.
\end{align}
The Discrete Fourier Transform, $\textrm{DFT}$, is obtained through the diagonalization of~$A_c$
\begin{align}\label{eqn:dspLambda-3}
    A_c {}&= \textrm{DFT}^H \Lambda \ \textrm{DFT},
\end{align}
where
\begin{align}\label{eqn:dspLambda-4}
    \Lambda {}&=\textrm{diag}\left[\lambda\right]\\
\label{eqn:dspLambda-5}
    \lambda {}&= \left[\lambda_0\cdots\lambda_{N-1}\right]^T=\left[1\ e^{-j\frac{2\pi}{N}}\cdots e^{-j\frac{2\pi}{N}(N-1)}\right]^T
    \\
    \label{eqn:dspLambda-6}
\textrm{DFT}{}&=\frac{1}{\sqrt{N}}\left[\hspace{-2mm}\begin{array}{llll}
1&1&\cdots&1\\
1&e^{-j\frac{2\pi}{N}}&\cdots&e^{-j\frac{2\pi}{N}(N-1)}\\
\vdots&\vdots&&\vdots\\
1&e^{-j\frac{2\pi}{N}(N-1)}&\cdots&e^{-j\frac{2\pi}{N}(N-1)(N-1)}
\end{array}\hspace{-2mm}\right]\\
\label{eqn:dspLambda-7}
{}&=\frac{1}{\sqrt{N}}\left[\lambda^0\cdots \lambda^{N-1}\right]\\
\label{eqn:dspLambda-8}
\textrm{DFT}^H{}&=\left[v_0\cdots v_{N-1}\right].
\end{align}
The matrix~$\Lambda$ in~\eqref{eqn:dspLambda-4} is the diagonal matrix of the eigenvalues, i.e., its diagonal entries are the graph frequencies. Equation~\eqref{eqn:dspLambda-5} defines the \textit{graph frequency vector} $\lambda$ and equation~\eqref{eqn:dspLambda-7} uses the notation  $\lambda^k=\lambda\odot\lambda\cdots\odot\lambda$ to represent $k$~times the Hadamard or entrywise product of the graph frequency vector~$\lambda$. By~\eqref{eqn:dspLambda-7} and~\eqref{eqn:dspLambda-8}, in DSP, the powers of the graph frequency vector $\lambda$  are conjugates of the eigenvectors, $\frac{1}{\sqrt{N}}\lambda^k=v_k^*$, and, by~\eqref{eqn:dspLambda-8}, the columns of the $\text{DFT}^H$ are the eigenvectors $v_k$ of~$A$.  The $\textrm{DFT}$ is symmetric, $\textrm{DFT}=\textrm{DFT}^T$, and unitary, $\text{DFT}^{-1} = \text{DFT}^H$.

\subsection{GSP basics}\label{subsec:gspbasics}
 We now let~$A$ be the adjacency matrix of an arbitrary (directed or undirected) graph~$G$ of~$N$ nodes and~$s$ be a graph signal. As observed for DSP and following \cite{Sandryhaila:13,Sandryhaila:14,Sandryhaila:14big}, in GSP, $A$ is the shift operator. It captures the local dependencies of the signal sample $s_n$ on the signal samples $s_m$ at the in-vertex neighbors $m\in \eta_n$  of~$n$ (given by the nonzero entries of row~$n$ of~$A$). The eigenvalues $\lambda_k$ and eigenvectors $v_k$ of~$A$ are the graph frequencies and graph spectral modes. Let graph frequency vector~$\lambda$ and matrix~$\Lambda$ be defined as before to collect the graph eigenvalues
\begin{align}\label{eqn:lambdaLambda-1}
\lambda {}&= [\lambda_0, \lambda_1, \hdots, \lambda_{N-1}]^T, \:\:\:
\Lambda = \textrm{diag}\left[\lambda\right].
\end{align}

The following assumptions hold even when not stated. On occasion, we state them explicitly.
\begin{assumption}[Strongly connected graph] \label{ass:connectedgraph}
The graph~$G$ is strongly connected.
\end{assumption}
Under assumption~\ref{ass:connectedgraph}, matrix~$A$ has no zero column or row.
\begin{assumption}[Distinct eigenfrequencies] \label{ass:eigunique}
The eigenvalues of $A$ are distinct.
\end{assumption}
Under assumption~\ref{ass:eigunique},\footnote{\label{ftn:nonderogatory} Distinct eigenvalues are assumed for simplicity. The results can be proved in the more general setting of~$A$ non-derogatory (equal minimum and characteristic polynomials (up to a factor $\pm1$), or, equivalently, the geometric multiplicity of any eigenvalue to be~1 (single eigenvector)).} $A$ is diagonalizable and the Graph Fourier Transform ($\textrm{GFT}$) is found\footnote{\label{ftn:Anotdiagonalizable-1} If assumption~\ref{ass:eigunique} does not hold, see \cite{derimoura-2017} for further details on the $\textrm{GFT}$.} by
\begin{align} \label{eqn:Aeigen}
	A = \textrm{GFT}^\textrm{-1} \ \Lambda \ \textrm{GFT}.
\end{align}
If~$A$ is symmetric, which is the case with undirected graphs, $\textrm{GFT}$ is orthogonal $\left(\textrm{GFT}^{-1}=\textrm{GFT}^T\right)$, and if~$A$ is normal then $\textrm{GFT}$ is unitary $\left(\textrm{GFT}^{-1}=\textrm{GFT}^H\right)$. For general graphs, $A$ is neither symmetric nor normal, but $\textrm{GFT}$ is full rank and invertible.\footnote{\label{ftn:computeinverseGFT} See \cite{domingosmoura-2020} for numerically stable diagonalization of~$A$ for directed graphs.} The spectral modes are the columns $v_k$ of $\textrm{GFT}^{-1}$.

The graph Fourier transform of graph signal~$s$ is
\begin{align}\label{eqn:graphFourier-1}
    \widehat{s}{}&=\textrm{GFT} s.
\end{align}
\begin{remark}[Fixing the \textrm{GFT}]\label{rmk:fixingGFT} It is well known \cite{gantmacher1959matrix} that the diagonalization of a matrix is unique up to reordering of the eigenvalues and normalization of the eigenvectors. Paralleling remark~\ref{rmk:Aequivalenceclass}, we assume the frequencies have been ordered and the eigenvectors appropriately normalized, see~\cite{shimoura-2021}, fixing the $\textrm{GFT}$ and~$\Lambda$.
\end{remark}
\textit{Cayley-Hamilton}. Let the characteristic polynomial of~$A$ be
\begin{align}\label{eqn:characteristicpoly-1}
\Delta(A){}&=c_0I+c_1A+\cdots+c_{N-1}A^{N-1}+A^N.
\end{align}
By the Cayley-Hamilton Theorem \cite{gantmacher1959matrix,lancaster1985theory,horn2012matrix}, $A$ satisfies its characteristic polynomial $\Delta(A)=0$ and so
\begin{align}\label{eqn:characteristicpoly-2}
    A^N=-c_0I-c_1A-\cdots-c_{N-1}A^{N-1},
\end{align}
 and $A^k$, $k\geq N$, is reduced by modular  arithmetic $\!\!\!\!\!\mod_{\!\!\Delta(A)}(\cdot)$.

\textit{Linear shift invariant (LSI) filtering}. Under assumption~\ref{ass:eigunique}, LSI filters in the vertex domain are polynomials $P(A)$ in the shift. By Cayley-Hamilton,\footnote{\label{ftn:characteriticminimalpoly-1} Under assumption~\ref{ass:eigunique}, the minimal polynomial of~$A$ equals $\Delta(A)$.}  $P(A)$ is at most degree~$N-1$.

The graph Fourier theorem \cite{Sandryhaila:13} parallels DSP's theorem
\begin{align}\label{eqn:FourierThem-1}
    P(A)\,s\xrightarrow{\mathcal{F}} P(\Lambda) \widehat{s},
\end{align}
and, in particular, the vertex shift relation
\begin{align}\label{eqn:vertexshift-1}
    A\,s\xrightarrow{\mathcal{F}} \Lambda\widehat{s}.
\end{align}

\textit{Spectral shift~$M$}. In \cite{shimoura-asilomar2019,shi2019graph,shimoura-2021}, we introduce a spectral graph shift~$M$ (see also \cite{leus2017dual} for a different definition) to shift a graph signal in the spectral domain preserving the dual of the shift invariance relation~\eqref{eqn:vertexshift-1}, i.e., such that
\begin{align}\label{eqn:spectralshift-1}
    \Lambda^*\,s\xrightarrow{\mathcal{F}} M\widehat{s}.
\end{align}
 References \cite{shimoura-asilomar2019,shi2019graph} show that
\begin{align}\label{eqn:spectralshift-2}
    M{}&=\textrm{GFT}\,\Lambda^*\textrm{GFT}^{-1},
\end{align}
and LSI spectral filters are polynomials $P(M)$.
\subsection{Graph impulse}\label{subsec:graphimpulse}
When studying graph signal representations, we need the concept of graph delta or graph impulse. In DSP, the impulse in the time domain and its Fourier transform are
\begin{align}\label{eqn:DSPdelta0-1}
    \delta_{t,0}{}&=e_0,\:\:\:\xrightarrow{\mathcal{F}}\:\:\: \widehat{\delta}_{t,0}=\textrm{DFT}\,\delta_{t,0}=\frac{1}{\sqrt{N}} 1,
\end{align}
where~$1$ is the vector of ones. In DSP, the time impulse $\delta_{t,0}$ is impulsive in the vertex domain (nonzero only at~$0$) and flat in the frequency domain. Further, the delayed time impulses $\delta_{t,n}=A_c^n\delta_{t,n}$ are centered at~$n$ and impulsive. Likewise, in DSP, the impulse in the frequency domain $\widehat{\delta}_{f,0}$ is impulsive now in frequency and flat in time. In other words, in DSP, the definition of impulse in time and frequency are symmetric\textemdash the time and frequency impulses are impulsive at $t=0$ and at $f=0$, respectively.

In GSP, in general, we either get impulsivity in one domain or flatness in the other, but not both. We have then two possible definitions for the vertex impulse and two possible definitions for the spectral impulse. We choose to preserve flatness and define delta graph signals that are flat in one domain. We discuss next how to define
\begin{inparaenum}[1)]
\item graph impulse signal in the vertex domain as the inverse GFT of a flat signal in the spectral domain; and
    \item graph impulse signal in the spectral domain as the GFT of a flat signal in the vertex domain.
        \end{inparaenum}
\subsubsection{Vertex graph impulse}\label{subsec:verteximpulse}
In the vertex domain, define the vertex impulse or delta $\delta_0$  as the inverse $\textrm{GFT}$ of a flat graph spectrum
\begin{align}\label{eqn:delta0}
\hspace{6.5mm}\delta_0{}&\xrightarrow{\mathcal{F}}\widehat{\delta}_0=\frac{1}{\sqrt{N}} 1\:\:\Longrightarrow\:\:
   \delta_0\delequal\textrm{GFT}^{-1}\left[\frac{1}{\sqrt{N}} 1\right].
\end{align}

The shifted replicas of the vertex graph impulse $\delta_0$ are
\begin{align}\label{eqn:delayedimpulsedeltan-a}
    \delta_n{}&=A^n\delta_0\xrightarrow{\mathcal{F}}\widehat{\delta}_n=\Lambda^n\frac{1}{\sqrt{N}}1=\frac{1}{\sqrt{N}}\lambda^n.
\end{align}
In GSP, the $\delta_n$'s, delayed $\delta_0$ by~$A^n$, are not impulsive.
\subsubsection{Spectral graph impulse}\label{subsubsec:spectralgraphimpulse}
We now consider the spectral graph impulse $\widehat{\delta}_{{\scriptsize\textrm{sp}},0}$ in the  spectral domain. We define it as the $\textrm{GFT}$ of a flat signal in the vertex domain
\begin{align}\label{eqn:deltasp0}
\delta_{{\scriptsize\textrm{sp}},0}=\frac{1}{\sqrt{N}}1{}&\xrightarrow{\mathcal{F}}\widehat{\delta}_{{\scriptsize\textrm{sp}},0}\:\:\Longrightarrow\:\:
   \widehat{\delta}_{{\scriptsize\textrm{sp}},0}\delequal\textrm{GFT}\left[\frac{1}{\sqrt{N}} 1\right].
\end{align}
The shifts of $\widehat{\delta}_{{\scriptsize\textrm{sp}},0}$ in the spectral domain are obtained with the spectral shift~$M$. Replicating~\eqref{eqn:delayedimpulsedeltan-a}, get
\begin{align}\label{eqn:spdeltashifts-1}
 \delta_{{\scriptsize\textrm{sp}},n}{}&=\Lambda^{*^{n}}\frac{1}{\sqrt{N}} 1=\frac{1}{\sqrt{N}}\lambda^{*^{n}}\xrightarrow{\mathcal{F}} \widehat{\delta}_{{\scriptsize\textrm{sp}},n}=M^n \widehat{\delta}_{{\scriptsize\textrm{sp}},0}.
\end{align}
\begin{remark}[Notation on vertex and spectral quantities]\label{rmk:vertexspectralquantities}
When referring to quantities using the shift~$A$ or the vertex impulse $\delta_0$ we will often not qualify them with the word ``vertex.'' In contrast, we will consistently qualify by ``spectral'' quantities related to the spectral shift~$M$ or the spectral impulse $\delta_{{\scriptsize\textrm{sp}},n}$ using the subscript `sp' as a reminder.
\end{remark}
\begin{remark}[Vertex and spectral impulses]\label{rmk:vertexspectralimpulses} We emphasize that we have two graph impulses, the \textit{vertex} impulse $\delta_0$ (that is flat in the spectral domain, see~\eqref{eqn:delta0}) and the \textit{spectral} impulse $\delta_{{\scriptsize\textrm{sp}},n}$ (that is flat in the vertex domain, see~\eqref{eqn:deltasp0}). In general, in GSP, neither is actually ``impulsive'' in either domain.
\end{remark}

\section{Vertex and Fourier Signal Representations}\label{sec:vertexgraphsignalrep}
At an abstract level, graph signals are vectors in an $N$~dimensional graph signal vector space $\mathbb{V}$ over field~$\mathbb{F}$. Amplifying, attenuating, adding, filtering, or processing signals is simplified by first expressing them as linear combinations of~$N$ basic signals. As a prelude to representations introduced in subsequent sections, here, we discuss first in subsection~\ref{subsec:graphsignalrep} a generic representation, and then in subsections~\ref{subsec:vertexEuclideanrep-1} and~\ref{subsec:spectralrep-1} consider the vertex and spectral representations, respectively.

\subsection{Graph signal representations}\label{subsec:graphsignalrep}
In the $N$-dimensional signal vector space~$\mathbb{V}$ over the field~$\mathbb{F}$, let $B_U=\left\{u_0,\cdots,u_{N-1}\right\}$ be a basis. Recall that the basis vectors $\left\{u_n\right\}_{0\leq n\leq N-1}$ are all nonzero and linearly independent. Mathematically, for any $s\in \mathbb{V}$:
\begin{tcolorbox}[ams align,height=3.5cm,valign=center,title={Graph signal representation},fonttitle=\bfseries\small]
\label{eqn:graphsignal-1A}
    s{}&=\left(s_U\right)_0u_0+\cdots+\left(s_U\right)_{N-1}u_{N-1}
    \\
    \label{eqn:graphsignal-2A}
    {}&=\underbrace{\left[u_0\cdots u_{N-1}\right]}_{U}\underbrace{\left[\begin{array}{c}
         \left(s_U\right)_0\\
         \vdots\\
         \left(s_U\right)_{N-1}
    \end{array}\right]}_{s_U}.
\end{tcolorbox}
\vspace{-2mm}
\begin{remark}[$\mathbb{V}\approx\mathbb{C}^N$]\label{rmk:VisomorphictoCN}
In the paper, we assume the field $\mathbb{F}=\mathbb{C}$, so, $s_U\in\mathbb{C}^N$. By~\eqref{eqn:graphsignal-1A}, $\mathbb{V}$ is isomorphic to  $\mathbb{C}^N$. In the sequel, we use this isomorphism and assume the signal space is the $N$-dimensional vector space $\mathbb{C}^N$ over the field $\mathbb{C}$.
\end{remark}
\begin{remark}[Ordered basis]\label{rmk:orderedbasis-1}
For $s_U$ to be well defined, the basis $B_U$ is ordered. If we reorder the basis by a permutation~$P$, the coordinate vector $s_U$ is itself reshuffled by~$P$:
\begin{align}\label{eqn:permutevertexordering-1}
s^P_U{}&=P\ s_U.
\end{align}
\end{remark}
By equation~\eqref{eqn:graphsignal-2A}, processing signals is computing with their coordinate vectors $s_U$. Because of its significance, this coordinatization of signals receives a special designation.
\begin{definition}[Representation]\label{def:representation-1} The \textit{representation} of~$s$ with respect to the \textbf{ordered} basis $B_U$ is its \textit{coordinate vector} $s_U$. The $n$th component $\left(s_U\right)_n$ of $s_U$ is the coefficient of the basis vector $u_n$ in the linear combination~\eqref{eqn:graphsignal-1A}.
\end{definition}
Choosing a signal representation corresponds to choosing a basis $B_U$. There are  infinitely many, with some particularly useful. DSP is essentially built around two representations, see section~\ref{sec:background}. For GSP, we consider six representations and discuss their specific advantages.

\subsection{Vertex, standard, or Euclidean representation}\label{subsec:vertexEuclideanrep-1}
The graph signal~$s$ is an indexed collection of samples $s=\left\{s_n\right\}_{n\in V}$, one at each vertex of the graph. The vertex graph signal representation is the natural one where the $n$th-component of the coordinate vector is the graph sample $s_n$ at indexing vertex $n\in V$ of the graph. This representation corresponds to the standard or Euclidean ordered basis $B_E=\left\{e_0,\cdots,e_{N-1}\right\}$. Clearly, $\left\{e_n\neq 0\right\}_{0\leq n\leq N-1}$ are linearly independent.  For easy reference, we formally present the vertex or Euclidean representation.
\begin{definition}[Vertex, standard, or Euclidean representation]\label{def:vertexrep}
The \textit{vertex, standard}, or \textit{Euclidean} representation of graph signal $s\in \mathbb{V}\approx\mathbb{C}^N$ is the coordinate vector of~$s$ with respect to the standard basis $B_E$.
\begin{tcolorbox}[ams align,height=2.8cm,valign=center,title={Vertex, standard, or Euclidean representation},fonttitle=\bfseries\small]
\label{eqn:euclideansignalrep-1}
s{}&=s_0e_0+\cdots+s_{N-1}e_{N-1}\\
\label{eqn:euclideansignalrep-2a}
{}&=\underbrace{\left[e_0\cdots e_{N-1}\right]}_{\scriptsize I_N}\left[
\begin{array}{c}
s_0\\
\vdots\\
s_{N-1}
\end{array}\right]=\left[
\begin{array}{c}
s_0\\
\vdots\\
s_{N-1}
\end{array}\right]=s_E.
\end{tcolorbox}
\end{definition}
Component~$n$ of the coordinate vector $s_E$ corresponds to~$n\in V$ and to~$e_n\in B_E$. Ordering $B_E$, orders nodes and  $s_E$ is well defined. Because the matrix with columns $e_n$ is the identity, we usually omit the subindex~$E$ and use the same symbol, e.g., $s$, for the graph signal~$s$ and its vertex representation $s_{E}$.

Reordering $B_E$ or the vertices permutes $s_E$ as in~\eqref{eqn:permutevertexordering-1}. In DSP, time is ordered and this issue is taken for granted. In GSP, to process signals, the ordering should be fixed and shared.
\subsection{Graph Fourier representation}\label{subsec:spectralrep-1}
Fourier analysis, frequency components, bandlimited, low pass come naturally from the spectral or Fourier transform domain description~$\widehat{s}$ of the signal~$s$. This can also be interpreted as a representation of~$s$ where the Fourier basis is:
\begin{align}\label{eqn:Fourierbasis-1}
    B_{\scriptsize\textrm{Fourier}}{}&=\left\{v_0,\cdots,v_{N-1}\right\},
\end{align}
where the eigenvectors $v_n$ of~$A$ are spectral modes and the columns of $\textrm{GFT}^{-1}$. We order the Fourier basis $B_{\scriptsize\textrm{Fourier}}$, ordering the spectral components $v_k$ and the graph frequencies $\lambda_k$.
The graph Fourier representation is formally presented next, again, for easy reference.
\begin{definition}[Graph Fourier  representation]\label{def:spectralrep}
The graph Fourier representation of~$s$ is its graph spectrum~$\widehat{s}$.
\begin{tcolorbox}[ams align,height=3.2cm,valign=center,title={Graph Fourier representation},fonttitle=\bfseries\small]
\label{eqn:graphspectralrep-1b}
s{}&=\widehat{s}_0v_0+\cdots+\widehat{s}_{N-1}v_{N-1}\\
\label{eqn:graphspectralrep-1c}
{}&=\underbrace{\left[\begin{array}{ccc}
v_0&\cdots& v_{N-1}
\end{array}\right]}_{\textrm{GFT}^{-1}}\underbrace{\left[\begin{array}{ccc}
\widehat{s}_0\\
\cdots\\
\widehat{s}_{N-1}
\end{array}\right]}_{\widehat{s}}
\end{tcolorbox}%
\end{definition}
%
\section{Vertex Impulsive Representation and Graph $z$-Transform}\label{sec:polynomialrep-a}
%
%
In this section, we consider several representations for the graph signal:
\begin{inparaenum}[1)]
\item as a linear combination of graph vertex impulses;
\item as the impulse response of a graph filter;
\item introduce the graph $z$-transform ($\textrm{G$z$T}$); and through the $\textrm{G$z$T}$ provide
\item a symbolic polynomial representation for graph signals.
\end{inparaenum}
\subsection{Vertex Impulsive Representation}\label{subsec:impulsiverep}
Consider the (ordered) set of the graph vertex impulse and its delayed replicas in~\eqref{eqn:delta0} and~\eqref{eqn:delayedimpulsedeltan-a}:
\begin{align}\label{eqn:impulsivebasis-1}
   B_{\scriptsize\textrm{imp}}{}&=\left\{\delta_0,\delta_1,\cdots,\delta_{N-1}\right\}= \left\{I\delta_0,A\delta_0,\cdots,A^{N-1}\delta_0\right\}.
\end{align}
To prove $B_{\scriptsize\textrm{imp}}$ is a basis, introduce the \textit{vertex impulsive} matrix $D_{\scriptsize\textrm{imp}}$ with columns the vectors in $B_{\scriptsize\textrm{imp}}$:
\begin{align}\label{eqn:impulsivebasis-4}
   D_{\scriptsize\textrm{imp}}{}&\delequal\left[\delta_0\,\delta_1\,\cdots\, \delta_{N-1}\right]=\left[A^0\delta_0\,A\delta_0\,\cdots\,A^{N-1}\delta_0\right].
\end{align}

We relate $D_{\scriptsize\textrm{imp}}$ to a Vandermonde matrix~$\mathcal{V}$.
\begin{result}[Vertex Impulsive and Vandermonde matrices]\label{res:impulsivevandermonde} 
\begin{align}\label{eqn:impulsivebasis-3}
    D_{\scriptsize\textup{imp}}\xrightarrow{\mathcal{F}}\frac{1}{\sqrt{N}}\mathcal{V},
\end{align}
where $\mathcal{V}$ is the Vandermonde matrix
\begin{align}\label{eqn:Vandermonde-1}
\hspace{-.25cm}\mathcal{V}{}&\hspace{-.1cm}=\hspace{-.1cm} \left[\hspace{-.15cm} \begin{array}{ccc}
   \lambda^0\hspace{-.15cm}&\hspace{-.15cm}\cdots\hspace{-.15cm}&\hspace{-.15cm}\lambda^{N-1}
   \end{array}\hspace{-.15cm} \right]
   \hspace{-.1cm}=\hspace{-.1cm}
   \left[\hspace{-.1cm} \begin{array}{lllll}
1&\lambda_0&\lambda_0^2&\cdots&\lambda_0^{N-1}\\
1&\lambda_1&\lambda_1^2&\cdots&\lambda_1^{N-1}\\
\vdots&\vdots&\vdots&\ddots & \vdots\\
1&\lambda_{N-1}&\lambda_{N-1}^2&\cdots&\lambda_{N-1}^{N-1}\\
\end{array}\hspace{-.1cm} \right].
\end{align}
\end{result}
\begin{proof}
This result follows by using~\eqref{eqn:delayedimpulsedeltan-a} for $\delta_n$ in $D_{\scriptsize\textup{imp}}$.
\end{proof}

\begin{result}[Full rank of vertex impulsive matrix]\label{res:impulsivefullrank} Under assumption~\ref{ass:eigunique},  $D_{\scriptsize\textup{imp}}$ is full rank.
    \end{result}
    \begin{proof} By result~\ref{res:impulsivevandermonde} and equation~\eqref{eqn:impulsivebasis-3}, $D_{\scriptsize\textrm{imp}}$ is the $\textrm{GFT}^{-1}$ of the Vandermonde matrix~$\mathcal{V}$. Under assumption~\ref{ass:eigunique},  $\mathcal{V}$ is full rank \cite{gantmacher1959matrix,lancaster1985theory,horn2012matrix}. Hence, $D_{\scriptsize\textrm{imp}}$ is full rank.
    \end{proof}
\begin{result}[Vertex impulsive basis]\label{res:impulsivebasis-a}
Under assumption~\ref{ass:eigunique}, $B_{\scriptsize\textup{imp}}$
is a basis\textemdash the vertex impulsive basis.
\end{result}
\begin{proof}
The vectors in $B_{\scriptsize\textrm{imp}}$ are the columns of  $D_{\scriptsize\textrm{imp}}$, which by result~\ref{res:impulsivefullrank} is full rank. Hence,  $B_{\scriptsize\textrm{imp}}$ is a basis.
\end{proof}

\begin{definition}[Vertex impulsive representation $p_{\scriptsize\textrm{imp}}$]\label{def:impulsiverep} The vertex impulsive representation of graph signal~$s$ is its coordinate vector $p_{\scriptsize\textrm{imp}}$ with respect to basis $B_{\scriptsize\textrm{imp}}$:
\begin{tcolorbox}[ams align,height=3.1cm,valign=center,title={Vertex impulsive representation},fonttitle=\bfseries\small]
\label{eqn:impulsiverep-a1}
    s{}&=p_0\delta_0+p_1\delta_1+\cdots+p_{N-1}\delta_{N-1}\\
    \label{eqn:impulsiverep-a2}
    &=\underbrace{\left[\begin{array}{cccc}
    \delta_0&\delta_1&\cdots&\delta_{N-1}
    \end{array}\right]}_{D_{\scriptsize\textrm{imp}}}\underbrace{\left[\begin{array}{c}
    p_0\\
    \cdots\\
    p_{N-1}
    \end{array}\right]}_{p_{\scriptsize\textrm{imp}}}
\end{tcolorbox}
\end{definition}

\textbf{Computing $p_{\scriptsize\textrm{imp}}$}. To find the coordinate vector $p_{\scriptsize\textrm{imp}}$ of~$s$  with respect to $B_{\scriptsize\textrm{imp}}$, in general, we solve  the linear system~\eqref{eqn:impulsiverep-a2}. In practice, a sparse approximation may suffice by minimizing $\left\|D_{\scriptsize\textrm{imp}}p_{\scriptsize\textrm{imp}}-s\right\|_2^2+\left\|p_{\scriptsize\textrm{imp}}\right\|_1$.
%

%
%
\subsection{Polynomial Transform Filter}\label{subsec:polytransformfilter}
Next, we interpret the vertex impulsive representation $p_{\scriptsize\textrm{imp}}$ as the coefficients of a linear shift invariant (LSI) graph filter. 
\begin{result}[$s$ as impulse response of $P_s(A)$] \label{res:simppolynomialfilter-1}
Let assumption~\ref{ass:eigunique} hold. Then the graph signal~$s$ is the impulse response 
\begin{align}\label{eqn:companionshift-1}
    s{}&=P_s(A)\delta_0
\end{align}
of the LSI polynomial filter
\begin{align}\label{eqn:sasimprespP(A)-1}
    P_s(A){}&=p_0I+p_1A+\cdots+p_{N-1}A^{N-1}
\end{align}
iff the vector of coefficients $p_{\scriptsize\textup{coef}}$ of $P_s(A)$ is the vertex impulsive representation $p_{\scriptsize\textup{imp}}$ in~\eqref{eqn:impulsiverep-a2}:
\begin{align}\label{eqn:sasimprespP(A)-3}
p_{\scriptsize\textup{coef}}{}=\left[\begin{array}{cccc}
p_0&
p_1&
\cdots&
p_{N-1}
\end{array}\right]^T=p_{\scriptsize\textup{imp}}.
\end{align}
\end{result}
\begin{proof}
The impulse response of the LSI $P_s(A)$ is
\begin{align}\label{eqn:sasimprespP(A)-2}
    P_s(A)\delta_0{}&=\left[p_0I+p_1A+\cdots+p_{N-1}A^{N-1}\right]\delta_0\\
\label{eqn:sasimprespP(A)-3a}    {}&=p_0I\delta_0+p_1A\delta_0+\cdots+p_{N-1}A^{N-1}\delta_0\\
\label{eqn:sasimprespP(A)-4}    {}&^=\underbrace{\left[\begin{array}{cccc}
\delta_0&\delta_1&\cdots&\delta_{N-1}
\end{array}\right]}_{D_{\scriptsize\textrm{imp}}}p_{\scriptsize\textup{coef}}.
\end{align}
Under assumption~\ref{ass:eigunique}, the impulse response of $P_s(A)$ is the graph signal~$s$ iff $p_{\scriptsize\textup{coef}}$ in~\eqref{eqn:sasimprespP(A)-4} equals $p_{\scriptsize\textup{imp}}$ in~\eqref{eqn:impulsiverep-a2}.
\end{proof}

\begin{definition}[Polynomial transform filter]\label{def:polynomialtransformfilter}
The LSI polynomial filter $P_s(A)$ in~\eqref{eqn:sasimprespP(A)-1} is the polynomial transform filter of~$s$.
\end{definition}

 The polynomial transform filter $P_s(A)$ in~\eqref{eqn:sasimprespP(A)-1} is in powers of~$A$. We provide an alternative description.
\begin{result}[Graph signal~$s$ and $P_s(A)$]\label{res:PsAands}
Given  $s\xleftrightarrow{\mathcal{F}}\widehat{s}$, its LSI polynomial transform filter $P_s(A)$ is alternatively given by
\begin{align}\label{eqn:PsAands-0}
P_s(A){}&=\textrm{GFT}^{-1}\textrm{diag}\left[\sqrt{N}\widehat{s}\right]\textrm{GFT}
\end{align}
\end{result}
\begin{proof}
  From~$s$ given as impulse response of~$P_s(A)$ in~\eqref{eqn:sasimprespP(A)-2}, using the diagonalization of $P_s(A)$, it successively follows
  \begin{align}\label{eqn:PsAimpulse-2b}
    P_s(A)\delta_0{}&=\textrm{GFT}^{-1}\cdot P_s\left(\Lambda\right)\cdot\textrm{GFT}\cdot\textrm{GFT}^{-1} \cdot\frac{1}{\sqrt{N}}1=s\\
  \label{eqn:PsAimpulse-2c}
  \Longrightarrow& P_s\left(\Lambda\right)\frac{1}{\sqrt{N}}1=\widehat{s}
  \Longrightarrow \frac{1}{\sqrt{N}}P_s\left(\Lambda\right)= \textrm{diag}\left[\widehat{s}\right]\\
    \label{eqn:PsAimpulse-2d}
  \Longrightarrow& P_s(A)=\textrm{GFT}^{-1}\textrm{diag}\left[\sqrt{N} \widehat{s}\right]\textrm{GFT},
  \end{align}
  where we used the definition of $\delta_0=\textrm{GFT}^{-1}\cdot\frac{1}{\sqrt{N}}1$.
\end{proof}

\subsection{Graph $z$-transform (G$z$T)}\label{subsec:graphztransform}
The powers of the shift~$A$ of the polynomial transform filter $P_s(A)$ in~\eqref{eqn:sasimprespP(A)-1} represent the GSP equivalent of the powers of the DSP shift $z^{-1}$. This motivates the following definition.
\begin{definition}[Graph $z$-transform (\textrm{G$z$T})]\label{def:graphpolynomialtransform} The \textrm{G$z$T} is
\begin{align}\label{eqn:graphpolynomialtransform}
    \textup{G$z$T}{}&=D_{\scriptsize\textup{imp}}^{-1}=\left[\begin{array}{cccc}
    \delta_0&\delta_1&\cdots&\delta_{N-1}
    \end{array}\right]^{-1}.
\end{align}
\end{definition}

In diagram form, the $\textrm{G$z$T}$ of~$s$ and its reconstruction are:
\begin{align}\label{eqn:simpisGPTofs-1}
s\xrightarrow{\hspace{3mm}\textup{G$z$T}\phantom{^{-1}}} p_{\scriptsize\textup{imp}}=\textup{G$z$T}\, s \xrightarrow{\hspace{3mm}\textrm{G$z$T}^{-1}} s
\end{align}
%
%
\begin{result}[$\textrm{G$z$T}^{-1}$ and $\mathcal{V}$]\label{res:GFTandGzT} The $\textrm{G$z$T}^{-1}$ and $\mathcal{V}$ are $\textrm{GFT}$ pairs:
\begin{align}\label{eqn:GPTandVandermonde-1}
    \textrm{G$z$T}^{-1}\xrightarrow{\textrm{GFT}} \frac{1}{\sqrt{N}}\mathcal{V}=\textrm{GFT}\cdot\textrm{G$z$T}^{-1}.
\end{align}
\end{result}
\begin{proof}
This follows from result~\ref{res:impulsivevandermonde} and definition~\ref{def:graphpolynomialtransform}.
\end{proof}

$\textrm{G$z$T}$ maps vertex signals~$s$ into $z$-transformed signals $p_{\scriptsize\textup{imp}}$.

The $\textrm{G$z$T}$ of~$s$ is the polynomial coefficient vector $p_{\scriptsize\textup{imp}}$ in~\eqref{eqn:sasimprespP(A)-3} that defines $P_s(A)$ in~\eqref{eqn:sasimprespP(A)-1}. To simplify notation, we introduce a symbolic polynomial representation $s(x)$.
\begin{definition}[Graph $z$-transform representation $s(x)$]\label{def:graphztransform} The graph $z$-transform representation of graph signal~$s$ is its coordinate vector $p_{\scriptsize\textrm{imp}}$ with respect to the monomial basis $B_{\scriptsize\textrm{monomial}}=\left\{1, x, \cdots, x^{N-1}\right\}$:
\begin{tcolorbox}[ams align,height=1.8cm,valign=center,title={Graph $z$-transform representation},fonttitle=\bfseries\small]
\label{eqn:GzTaspolys(x)-2}
\textrm{G$z$T}\,s\approx s(x){}&=p_0+p_1\,x+p_2\,x^2+\cdots+p_{N-1}\,x^{N-1}\\
\label{eqn:GzTaspolys(x)}
{}&=\left[\begin{array}{ccccc}1&x&x^2&\cdots&\end{array} x^{N-1}\right]p_{\scriptsize\textup{imp}}.
\end{tcolorbox}
\end{definition}
\begin{remark}[Various meanings for $p_{\scriptsize\textup{imp}}$]\label{rmk:imterpretationpimp}
We have multiple interpretations for $p_{\scriptsize\textup{imp}}$:
\begin{inparaenum}[1)]
\item as coordinate vector of~$s$ with respect to basis $B_{\scriptsize\textup{imp}}$;
    \item determining the polynomial transform filter $P_s(A)$;
    \item as \textrm{G$z$T} of~$s$; and
    \item defining $s(x)$.
\end{inparaenum}
We will take advantage of these several understandings.
\end{remark}
\subsection{Spectral Impulsive Representation}\label{subsec:spectralimpulsiverep}
Section~\ref{sec:vertexgraphsignalrep} discusses the standard or Euclidean representation of graph signals~$s$ (with respect to basis $B_E$), as well as the Fourier representation (with respect to basis $B_{\scriptsize\textrm{Fourier}}$). Section~\ref{sec:polynomialrep-a} presents the representation of graph signals with respect to the basis $B_{\scriptsize\textrm{imp}}$ whose basis vectors are the vertex impulse $\delta_0$ and its delayed replicas defined in section~\ref{subsec:verteximpulse}.
 In section~\ref{subsubsec:spectralgraphimpulse} and equation~\eqref{eqn:deltasp0}, we define the graph impulse $\widehat{\delta}_{{\scriptsize\textrm{sp}},0}$, but now in the spectral domain. As~\eqref{eqn:deltasp0} shows, $\delta_{{\scriptsize\textrm{sp}},0}$, the $\textrm{GFT}^{-1}$ of the spectral graph impulse, is flat in the vertex domain. This section considers the representation of graph signal~$s$ with respect to $\delta_{{\scriptsize\textrm{sp}},0}$ and its delayed replicas.

 Consider the set of $\delta_{{\scriptsize\textrm{sp}},0}$ and its spectral shifts:
 \begin{align}\label{eqn:graphfrequencyrepresentation-1a}
    B_{\scriptsize\textrm{sp,imp}}{}&=\left\{\delta_{{\scriptsize\textrm{sp}},0}, \delta_{{\scriptsize\textrm{sp}},1},\cdots, \delta_{{\scriptsize\textrm{sp}},{N-1}}\right\}.
\end{align}
Collect the vectors in $B_{\scriptsize\textrm{sp,imp}}$ in the spectral impulse matrix\footnote{\label{rmk:spectralimpulsematrix} Note that the columns of $D_{\scriptsize\textrm{sp,imp}}$ are flat, not impulsive.}
\begin{align}\label{eqn:Dspimp-1}
D_{\scriptsize\textrm{sp,imp}}{}&=\left[\begin{array}{cccc}
\delta_{{\scriptsize\textrm{sp}},0}& \delta_{{\scriptsize\textrm{sp}},1}&\cdots& \delta_{{\scriptsize\textrm{sp}},{N-1}}
\end{array}\right].
\end{align}
\begin{result}[$D_{\scriptsize\textrm{sp,imp}}$ and $\mathcal{V}^*$]\label{res:DspimpandV*}
\begin{align}\label{eqn:graphfrequencyrepresentation-1}
D_{\scriptsize\textrm{sp,imp}}{}&=\frac{1}{\sqrt{N}}\mathcal{V}^*
=\frac{1}{\sqrt{N}}\left[\begin{array}{cccc}1& \lambda^{*}&\cdots& \lambda^{*{^{N-1}}}\end{array}\right].
\end{align}
\end{result}
\begin{proof}
Result follows from equation~\eqref{eqn:spdeltashifts-1}.
\end{proof}

    Equation~\eqref{eqn:graphfrequencyrepresentation-1} shows that the vectors of the set $B_{\scriptsize\textrm{sp,imp}}$ are, apart a scaling, the columns of~$\mathcal{V}^*$.

\begin{result}[Spectral impulse basis $B_{\scriptsize\textrm{sp,imp}}$]\label{res:graphfreqbasis-a}
Under assumption~\ref{ass:eigunique}, $B_{\scriptsize\textrm{sp,imp}}$
is a basis.
\end{result}
\begin{proof}
   By assumption~\ref{ass:eigunique}, $\mathcal{V}^*$ is full rank.
\end{proof}

\begin{definition}[Spectral impulsive representation]\label{def:spectralimpulserepresentation}
The spectral impulsive representation of graph signal~$s$ is its coordinate vector $q_{\scriptsize\textrm{sp,imp}}$ with respect to basis $B_{\scriptsize\textrm{sp,imp}}$:
  \begin{tcolorbox}[ams align,height=3.6cm,valign=center,title={Spectral impulsive representation},fonttitle=\bfseries\small]
\label{eqn:sgraphfreqvectorrep-1}
 \hspace{-4mm}  s{}&=q_0\delta_{{\scriptsize\textrm{sp}},0}+q_1\delta_{{\scriptsize\textrm{sp}},1}+\cdots+q_{N-1}\delta_{{\scriptsize\textrm{sp}},{N-1}}\\
    &=D_{\textrm{sp,imp}}\underbrace{\left[\hspace{-1mm}\begin{array}{c}
    q_0\\
    q_1\\
    \cdots\\
    q_{N-1}
   \end{array}\hspace{-1mm}\right]}_{q_{\scriptsize\textrm{sp,imp}}}
   \label{eqn:sgraphfreqvectorrep-2}
   =\frac{1}{\sqrt{N}}\mathcal{V}^*q_{\scriptsize\textrm{sp,imp}}
\end{tcolorbox}
\end{definition}
\section{Companion Model\textemdash a Canonical GSP Model}\label{sec:GSPcompanionmodel}
In DSP, the DSP cyclic shift~$A$ in~\eqref{eqn:Adsp} acts on graph signal~$s$ as given in~\eqref{eqn:periodicshift-1}. Decompose the cyclic shift as
\begin{align} \label{eqn:Adsp-bc}
    A_c {}&= \underbrace{\begin{bmatrix}
	0  & 0  & \hdots & 0 & 0 \\
	1  & 0  &  \hdots & 0 & 0 \\
	\vdots  & 1 & \ddots & \vdots  & \vdots\\
	\vdots  & \vdots & \ddots & \ddots  & \vdots\\
	0  & 0 &  \hdots & 1 & 0 \\
    \end{bmatrix}}_{{A_{c,\textrm{line shift}}}}+\underbrace{\begin{bmatrix}
	0  & 0  & \hdots & 0 & 1 \\
	0 & 0  &  \hdots & 0 & 0 \\
	\vdots  & 0 & \ddots & \vdots  & \vdots\\
	\vdots  & \vdots & \ddots & \ddots  & \vdots\\
	0  & 0 &  \hdots & 0 & 0 \\
    \end{bmatrix}}_{{A_{c,\textrm{periodic bc}}}}.
\end{align}
Then the shifted time signal $As$ is delayed (moved downwards) by the \textit{line shift} (left block in~\eqref{eqn:Adsp-bc}) and the signal extension $s_N$ is determined by the periodic boundary condition (right block in~\eqref{eqn:Adsp-bc}) \cite{Pueschel:05e,Pueschel:08a,Pueschel:08b} that wraps around the time signal so that sample $s_{N-1}$ reappears as the first component of $As$.

In this section, we look for a GSP signal model where the graph shift acts in similar fashion to~\eqref{eqn:Adsp-bc}. We accomplish it with the \textit{impulsive} GSP signal representation. The resulting GSP model leads to the \textit{companion} shift and the \textit{companion} graph. These are \textit{canonical} shift and \textit{canonical} graph representations to which, under assumptions~\ref{ass:connectedgraph} and~\ref{ass:eigunique}, every other generic GSP model can be reduced to. We use these and the $\textrm{G$z$T}$ of section~\ref{subsec:graphztransform}, to introduce a \textit{fast graph convolution} in section~\ref{sec:graphconvolution} .
\subsection{Canonical Companion Shift}\label{subsec:canonialcshift}
To obtain the representation of~$A$ with respect to $B_{\scriptsize\textup{imp}}$, we apply the shift to each vector $\delta_n\in B_{\scriptsize\textup{imp}}$. Get
\begin{align}\label{eqn:getcomapnionshift-1}
  A\delta_0 {}& \!=\!\delta_1,\! \cdots\!, A\delta_n \! =\!A^{n+1}\delta_0\!=\!\delta_{n+1}, \!\cdots\!,  A\delta_{N-2}\!=\!\delta_{N-1}\!.
\end{align}
We need a ``signal extension'' or ``boundary condition'' for
\begin{align}\label{eqn:getcomapnionshift-2}
  A\delta_{N-1} {}& =A^{N}\delta_0.
\end{align}
This boundary condition is embedded in the matrix~$A$ and is obtained by reducing it by Cayley-Hamilton Theorem. Applying then this theorem through equation~\eqref{eqn:characteristicpoly-2},
\begin{align}\label{eqn:companionshift-5}
\hspace{-4mm}  A\delta_{N-1} {}& =-c_0I\delta_0\!-\!c_1A \delta_0\!-\!c_2A^2 \delta_0 -\!\cdots\!-\!c_{N-1}A^{N-1}\delta_0.
\end{align}
The boundary condition in~\eqref{eqn:companionshift-5} for $A\delta_{N-1}$ is a linear combination of the basis vectors $A\delta_n\in B_{\scriptsize\textup{imp}}$. The coefficients of the linear combination are the negative of the coefficients $c_n$ of the characteristic polynomial $\Delta_A(x)$ of~$A$ given in~\eqref{eqn:characteristicpoly-1}.

Putting together the~$N$ equations~\eqref{eqn:getcomapnionshift-1}-\eqref{eqn:getcomapnionshift-2} and using the boundary condition~\eqref{eqn:companionshift-5},
\begin{align}\label{eqn:companionshift-9a}
\!\!\!\!\!\!\! A&\left[\begin{array}{cccc}
\delta_0&\delta_1&\cdots&\delta_{N-1}
\end{array}\right] =
\\
\label{eqn:companionshift-9}
{}
\!\!\!\!\!\!\!=&\left[\begin{array}{cccc}
\delta_0&\delta_1&\cdots&\delta_{N-1}
\end{array}\right]\!\!
\underbrace{\left[\!\!\begin{array}{ccccl}
0&0&\cdots&0&-c_0\\
1&0&\cdots&0&-c_1\\
0&1&\ddots&0&-c_2\\
\vdots&\vdots&\ddots&\ddots&\vdots\\
0&0&\cdots&1&-c_{N-1}\\
\end{array}\!\!\!\right]}_{C_{\textrm{comp}}}\!.
\end{align}
Equation~\eqref{eqn:companionshift-9} shows that the representation of the shift with respect to $B_{\scriptsize\textup{imp}}$ is the \textit{companion matrix}  $C_{\textrm{comp}}$. It is the \textit{companion} matrix \cite{gantmacher1959matrix,lancaster1985theory,horn2012matrix} of the characteristic polynomial $\Delta_A(x)$ of the graph shift~$A$. We refer to $C_{\textrm{comp}}$ as the \textit{companion shift}. We can rewrite it as:
\begin{align}
\label{eqn:companionshift-7}
\hspace{-4mm}C_{\textrm{comp}}{}&\hspace{-1mm}=\hspace{-1mm}
\underbrace{\left[\hspace{-1mm}\begin{array}{ccccc}
0\hspace{-0.5mm}&\hspace{-0.5mm}0\hspace{-0.5mm}&\hspace{-0.5mm}\cdots\hspace{-0.5mm}&\hspace{-0.5mm}0\hspace{-0.5mm}&\hspace{-0.5mm}1\\
1\hspace{-0.5mm}&\hspace{-0.5mm}0\hspace{-0.5mm}&\hspace{-0.5mm}\cdots\hspace{-0.5mm}&\hspace{-0.5mm}0\hspace{-0.5mm}&\hspace{-0.5mm}0\\
0\hspace{-0.5mm}&\hspace{-0.5mm}1\hspace{-0.5mm}&\hspace{-0.5mm}\ddots\hspace{-0.5mm}&\hspace{-0.5mm}0\hspace{-0.5mm}&\hspace{-0.5mm}0\\
\vdots\hspace{-0.5mm}&\hspace{-0.5mm}\vdots\hspace{-0.5mm}&\hspace{-0.5mm}\ddots\hspace{-0.5mm}&\hspace{-0.5mm}\ddots\hspace{-0.5mm}&\hspace{-0.5mm}\vdots\\
0\hspace{-0.5mm}&\hspace{-0.5mm}0\hspace{-0.5mm}&\hspace{-0.5mm}\cdots\hspace{-0.5mm}&\hspace{-0.5mm}1\hspace{-0.5mm}&\hspace{-0.5mm}0\\
\end{array}\hspace{-1mm}\right]}_{A_c}
\!+\!
%
\underbrace{\left[\hspace{-1mm}\begin{array}{c}
-1-c_0\\
-c_1\\
\cdots\\
-c_{N-1}\\
\end{array}\hspace{-2mm}\right]
\left[0\,0\ \cdots\ 0\ 1\right]}_{\textrm{rank 1}}.
\end{align}
Equation~\eqref{eqn:companionshift-7} gives $C_{\textrm{comp}}$ as the sum of a unitary matrix, the DSP cyclic shift $A_c$, plus a rank one matrix. On the other hand, we may decompose $C_{\textrm{comp}}$ as:
\begin{align}
\label{eqn:Agsp-bc}
\hspace{-4mm}C_{\textrm{comp}}{}&=
\underbrace{\left[\hspace{-1mm}\begin{array}{ccccc}
0\hspace{-0.5mm}&\hspace{-0.5mm}0\hspace{-0.5mm}&\hspace{-0.5mm}\cdots\hspace{-0.5mm}&\hspace{-0.5mm}0\hspace{-0.5mm}&\hspace{-0.5mm}0\\
1\hspace{-0.5mm}&\hspace{-0.5mm}0\hspace{-0.5mm}&\hspace{-0.5mm}\cdots\hspace{-0.5mm}&\hspace{-0.5mm}0\hspace{-0.5mm}&\hspace{-0.5mm}0\\
0\hspace{-0.5mm}&\hspace{-0.5mm}1\hspace{-0.5mm}&\hspace{-0.5mm}\ddots\hspace{-0.5mm}&\hspace{-0.5mm}0\hspace{-0.5mm}&\hspace{-0.5mm}0\\
\vdots\hspace{-0.5mm}&\hspace{-0.5mm}\vdots\hspace{-0.5mm}&\hspace{-0.5mm}\ddots\hspace{-0.5mm}&\hspace{-0.5mm}\ddots\hspace{-0.5mm}&\hspace{-0.5mm}\vdots\\
0\hspace{-0.5mm}&\hspace{-0.5mm}0\hspace{-0.5mm}&\hspace{-0.5mm}\cdots\hspace{-0.5mm}&\hspace{-0.5mm}1\hspace{-0.5mm}&\hspace{-0.5mm}0\\
\end{array}\hspace{-1mm}\right]}_{C_{\textrm{line shift}}}
+
%
\underbrace{\left[\hspace{-1mm}\begin{array}{ccccl}
0\hspace{-1mm}&\hspace{-1mm}0\hspace{-1mm}&\hspace{-1mm}\cdots\hspace{-1mm}&\hspace{-1mm}0&\hspace{-1mm}-c_0\\[-2pt]
0\hspace{-1mm}&\hspace{-1mm}0\hspace{-1mm}&\hspace{-1mm}\cdots\hspace{-1mm}&\hspace{-1mm}0\hspace{-1mm}&\hspace{-1mm}-c_1\\[-2pt]
0\hspace{-1mm}&\hspace{-1mm}0\hspace{-1mm}&\hspace{-1mm}\cdots\hspace{-1mm}&\hspace{-1mm}0\hspace{-1mm}&\hspace{-1mm}-c_2\\[-2pt]
\vdots\hspace{-1mm}&\hspace{-1mm}\vdots\hspace{-1mm}&\hspace{-1mm}\ddots\hspace{-1mm}&\hspace{-1mm}\ddots\hspace{-1mm}&\hspace{-1mm}\vdots\\[-2pt]
0\hspace{-1mm}&\hspace{-1mm}0\hspace{-1mm}&\hspace{-1mm}\cdots\hspace{-1mm}&\hspace{-1mm}0\hspace{-1mm}&\hspace{-1mm}-c_{N-1}
\end{array}\hspace{-2mm}\right]}_{C_{\textrm{linear bc}}}.
\end{align}
Equation~\eqref{eqn:Agsp-bc} resolves $C_{\textrm{comp}}$ as a `line shift' $C_{\textrm{line shift}}$ corrected by a `boundary condition'  $C_{\textrm{linear bc}}$. It replicates the structure of the DSP cyclic shift given in~\eqref{eqn:Adsp-bc}. Like $A_{c,\textrm{line shift}}$ in~\eqref{eqn:Adsp-bc},  $C_{\textrm{line shift}}$ moves the graph signal downwards, while $C_{\textrm{linear bc}}$ retains the coefficients $\left\{-c_n\right\}_{0\leq n\leq N-1}$ of the boundary condition. This is a more general boundary condition than for the cyclic shift, since for $A_{c,\textrm{periodic bc}}$ all $c_n=0$, except $c_0=-1$, see~\eqref{eqn:Adsp-bc}. This agrees with the characteristic polynomial of the DSP cyclic shift of for which $\Delta_{A_{\textrm{c}}}\left(x\right)=x^N-1$.

Since  $C_{\textrm{comp}}$ is determined by the characteristic polynomial $\Delta_A(x)$, it only depends on the graph frequencies or eigenvalues of~$A$, not on the spectral modes or eigenvectors of~$A$. And this shows that, under diagonalization of~$A$, we can associate to arbitrary adjacency matrices a \textit{canonical} weighted adjacency matrix, its \textit{companion} shift.

\begin{result}[Diagonalization of $C_{\textrm{comp}}$]\label{res:diagonalizationcompanion} Under assumption~\ref{ass:eigunique}, $C_{\textrm{comp}}$ is diagonalized by the Vandermonde matrix
\begin{align}\label{eqn:companiondiagonalization}
C_{\textrm{comp}}=\mathcal{V}^{-1}\Lambda\ \mathcal{V}.
\end{align}
\end{result}
\begin{proof} This is a well known result. It can be verified by direct substitution that $\left[1\,\lambda_i\,\cdots\,\lambda_i^{N-1}\right]$ is a left eigenvector of $C_{\textrm{comp}}$ for eigenvalue $\lambda_i$, from which the result follows.
\end{proof}

\textit{Companion graph Fourier transform}. Given~\eqref{eqn:companiondiagonalization}, the Vandermonde matrix $\mathcal{V}$ is the graph Fourier transform for signals in impulsive representation, replicating the DSP result where the $\textrm{DFT}$ is the Vandermonde matrix of the eigenfrequencies (apart a normalizing factor), see~\eqref{eqn:dspLambda-6}.

This shows that the impulsive representation replicates for GSP another dimension of DSP. In fact, just like for DSP, the eigenvalues (frequencies) provide the whole picture, since the companion graph Fourier transform is defined by the frequency vectors $\lambda$ and its powers.

Next, we associate a weighted \textit{companion} graph $G_{\textit{comp}}$ to $C_{\textrm{comp}}$. Both of these, $G_{\textit{comp}}$ and $C_{\textrm{comp}}$, are \textit{canonical} graph representations connected with any GSP graph.
\subsection{Canonical Companion Graph}\label{subsec:companiongraph}
\begin{figure}[htb!]
\centering	\includegraphics[width=8cm, keepaspectratio]{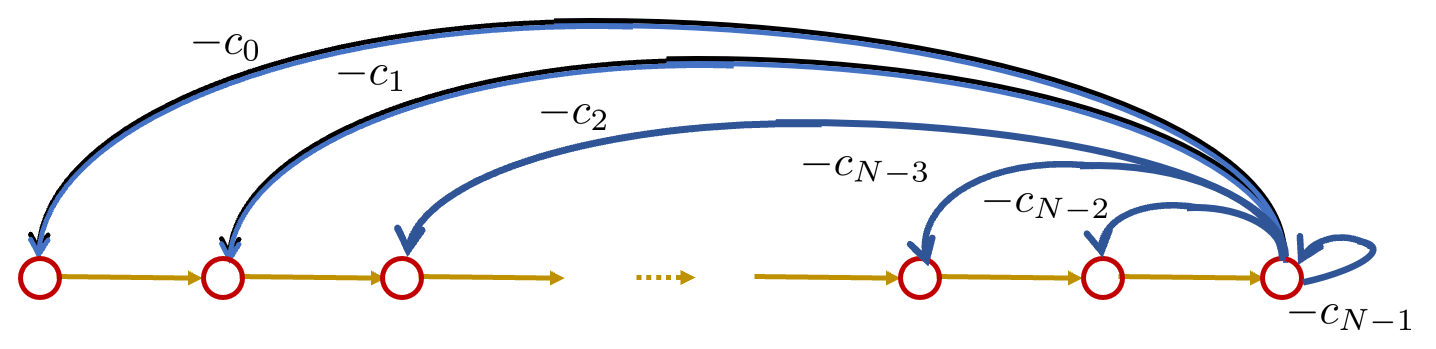}
	\caption{Companion graph. Unlabelled edges have weight~1. Other edges labeled by their weights.}
		\label{fig:Sgraph}
\end{figure}

 The companion matrix $C_{\textrm{comp}}$ defines the (weighted) companion graph $G_{\textrm{comp}}=\left(V_{\textrm{comp}}, E_{\textrm{comp}}\right)$ displayed in figure~\ref{fig:Sgraph}. Under assumption~\ref{ass:eigunique}, any directed or undirected signal graph~$G$ has a corresponding weighted \textit{companion} graph.

 The structure of the companion graph in figure~\ref{fig:Sgraph} extends the structure of the DSP cyclic graph in figure~\ref{fig:directedcyclegraph}. The DSP cyclic graph  follows form the companion graph of figure~\ref{fig:Sgraph} by taking $c_0=-1$ and eliminating the self-loop and all the remaining backward pointing edges.

The \textit{companion} graph $G_{\textrm{comp}}$ has a canonical structure:
\begin{inparaenum}[1)]
\item its node set $V_{\textrm{comp}}$ has~$N$ nodes, node~$n$ is associated with basis vector $\delta_n\in B_{\textrm{imp}}$ (or power $A^n$). In other words, these nodes are not the nodes of the original graph~$G$ associated with~$A$;
    \item it is directed;
        \item the edge set $E_{\textrm{comp}}$ combines a directed path graph with possibly a self-loop at node $N-1$ and up to~$N-1$ directed backward edges pointing from node~$N-1$ to the previous nodes;
            \item these directed edges are weighted by the negative of the coefficient $c_n$ of $\Delta_A(x)$;
                \item iff $c_0\neq 0$, the companion graph is strongly connected. This is the case if zero is not an eigenvalue of~$A$.
                    \end{inparaenum}

\subsection{Example}
\begin{figure}[htb!]
\centering	\includegraphics[width=6cm, keepaspectratio]{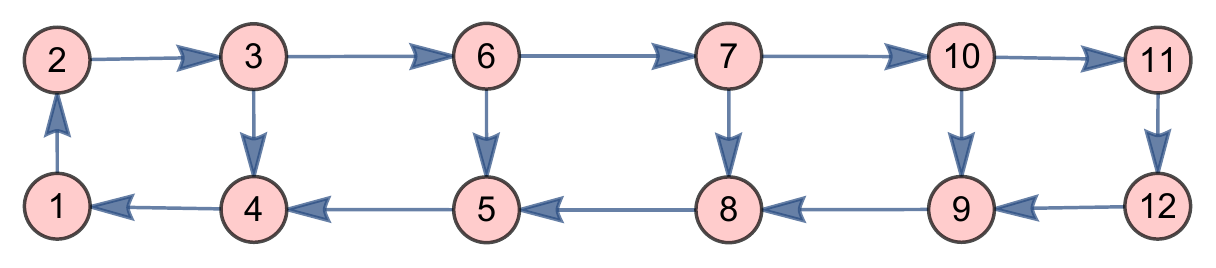}\\
\centering	\includegraphics[width=6cm, keepaspectratio]{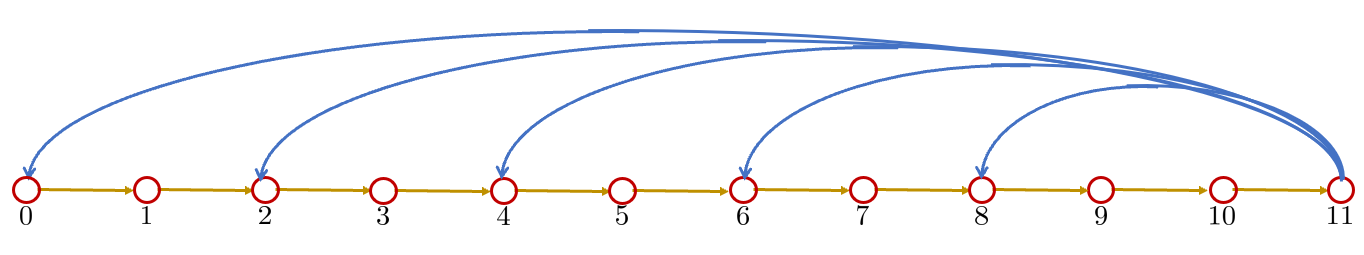}
	\caption{Directed ladder graph and its companion graph.}
		\label{fig:ladderdir12}
\end{figure}

\begin{figure}[htb!]
\centering	\includegraphics[width=5cm, keepaspectratio]{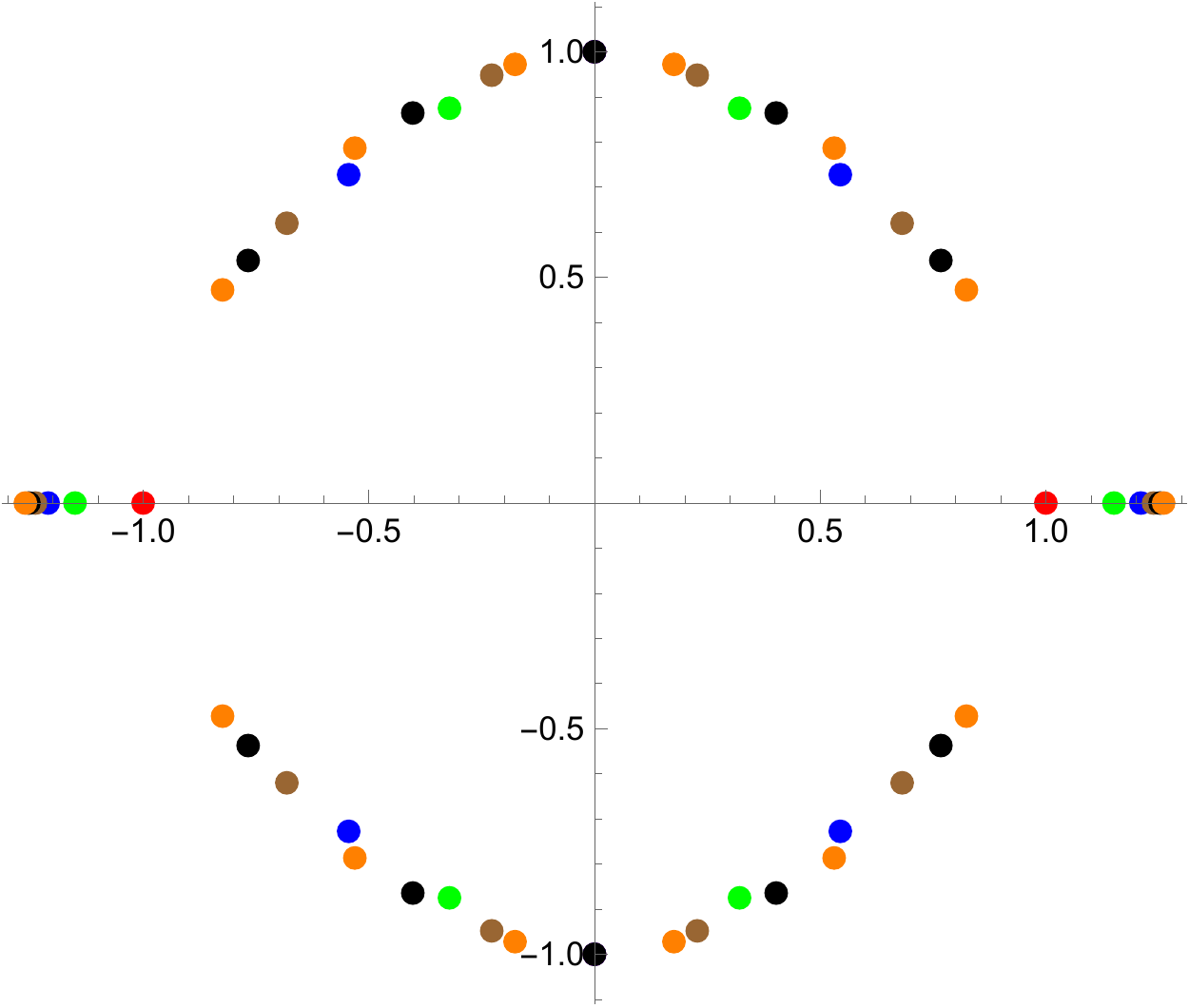}\\
	\caption{Eigenfrequencies of directed ladder graphs with 4 (red), 6 (green), 8 (blue), 10 (brown), 12 (black), and 14 (orange) nodes.}
	\label{fig:eigenfrequenciesladderdir04-14}
\end{figure}

Figure~\ref{fig:ladderdir12} shows on top a ``directed'' ladder graph with 12 nodes and below it the corresponding canonical graph. The characteristic polynomial of the adjacency matrix of a ladder graph like shown in the figure but with $2k$ nodes is
\begin{align}\label{eqn:Deltaladderdirected12}
\Delta_A(x){}&=-1-x^2-x^4-x^8-\cdots-x^{2(k-2)}+x^{2k}.
\end{align}
The polynomial $\Delta_A(x)$ explains why the edge weights of the companion graph of the directed ladder graph are all ones (the coefficients of $\Delta_A(x)$ are  $c_n\equiv-1$).
The eigenfrequencies of this directed ladder graph are illustrated for k=4,6,8,10,12, and 14 nodes in figure~\ref{fig:eigenfrequenciesladderdir04-14}. They distribute close to the unit circle.

\begin{figure}[htb!]
\centering	\includegraphics[width=4cm, keepaspectratio]{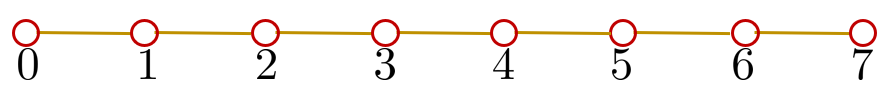}\hspace{3mm}
\includegraphics[width=4cm, keepaspectratio]{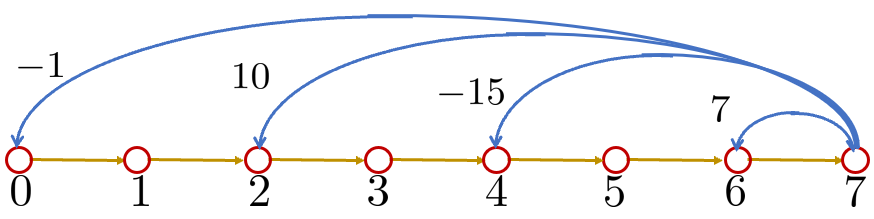}
	\caption{Path graph and its canonical companion graph.}
	\label{fig:undirpath8}
\end{figure}

As another example, consider the undirected $N$~node path. Its characteristic polynomial is the 3-term recursion
\begin{align}\label{eqn:charpolyundirpathNnodes}
\Delta_N(x){}&=x\Delta_{N-1}(x)-\Delta_{N-2}(x),\,\, \Delta_0(x)=1, \Delta_1(x)=\frac{x}{2}.
\end{align}
This gives $\Delta_N(x)=U\left(\frac{x}{2}\right)$ where $U(x)$ is the Chebyshev polynomial of the second kind \cite{mason2002chebyshev}. For example, for $N=8$
\begin{align}\label{eqn:charpolyundirpath}
\Delta_8(x){}&=x^8-7x^6+15x^4-10x^2+1
\end{align}
The path and its companion graph are in figure~\ref{fig:undirpath8}.
%
%
\section{Representations for the spectrum~$\widehat{s}$}\label{sec:repwidehats}
Taking the $\textrm{GFT}$ to both sides of a representation of~$s$, we obtain a representation for~$\widehat{s}$. We consider two representations for~$\widehat{s}$, derived from the vertex and the spectral impulsive representations given by~\eqref{eqn:impulsiverep-a1} and~\eqref{eqn:sgraphfreqvectorrep-1}, respectively.

\subsection{Spectrum vertex impulse representation}\label{subsec:spectrumverteximpulsiverep}
 Take the $\textrm{GFT}$ to both sides of~\eqref{eqn:impulsiverep-a1}. Using result~\ref{res:impulsivevandermonde} and equation~\eqref{eqn:delayedimpulsedeltan-a}, obtain the representation of~$\widehat{s}$ with respect to
 \begin{align}\label{eqn:frqvectorbasis-1}
B_\lambda{}&=\left\{\frac{1}{\sqrt{N}}1,\frac{1}{\sqrt{N}}\lambda,\cdots, \frac{1}{\sqrt{N}}\lambda^{N-1}\right\}.
\end{align}
The set $B_\lambda$ is a basis. In fact, its vectors, apart the scaling factor $\frac{1}{\sqrt{N}}$, are the columns of the Vandermonde matrix~$\mathcal{V}$, and~$\mathcal{V}$ is full rank under assumption~\ref{ass:eigunique}.
\begin{definition}[Spectrum vertex impulsive representation]\label{def:graphspfreqvecrep} The spectrum vertex impulsive representation of $\widehat{s}=\textrm{GFT}\ s$ is the coordinate vector $p_{\scriptsize\textrm{imp}}$ with respect to basis $B_{\lambda}$.
\begin{tcolorbox}[ams align,height=3.1cm,valign=center,title={Spectrum vertex impulsive representation},fonttitle=\bfseries\small]
\label{eqn:shatverteximpulsiverep-1}
\hspace{-3mm}    \widehat{s}{}&=p_0\widehat{\delta}_0+p_1\widehat{\delta}_1+\cdots+p_{N-1}\widehat{\delta}_{N-1}
    \\
\label{eqn:shatverteximpulsiverep-3}
   {}&= \underbrace{\left[\hspace{-2mm} \begin{array}{cccc}
    \frac{1}{\sqrt{N}}\lambda^0\hspace{-2mm}&\frac{1}{\sqrt{N}}\lambda\hspace{-2mm}&\cdots\hspace{-2mm}&\frac{1}{\sqrt{N}}\lambda^{N-1}
    \end{array}\hspace{-2mm}\right]}_{\frac{1}{\sqrt{N}}\mathcal{V}}\underbrace{\left[\begin{array}{c}
    p_0\\
    \cdots\\
    p_{N-1}
    \end{array}\right]}_{p_{\scriptsize\textrm{imp}}}
\end{tcolorbox}
\end{definition}
The spectrum vertex impulsive representation for~$\widehat{s}$ has the same coordinate vector $p_{\scriptsize\textrm{imp}}$ as the vertex impulsive representation for~$s$. It is the basis that is different. Now, it is the frequency vector $\lambda$ and its powers that are the basis vectors for this representation of~$\widehat{s}$

\subsection{Spectrum spectral impulse representation}\label{subsec:spectrumspimpulsiverep}
Taking the $\textrm{GFT}$ to both sides of equation~\eqref{eqn:sgraphfreqvectorrep-1}, we obtain the representation of $\widehat{s}=\textrm{GFT}\ s$ with respect to
\begin{align}
\label{eqn:graphspimpulsebasis-2}
\widehat{B}_{\scriptsize\textrm{sp,imp}}{}&=\left\{\widehat{\delta}_{{\scriptsize\textrm{sp}},0}, \widehat{\delta}_{{\scriptsize\textrm{sp}},1},\cdots, \widehat{\delta}_{{\scriptsize\textrm{sp}},{N-1}}\right\}\\
\label{eqn:graphspimpulsebasis-2b}
{}&=\left\{M^0\widehat{\delta}_{{\scriptsize\textrm{sp}},0}, M\widehat{\delta}_{{\scriptsize\textrm{sp}},0},\cdots, M^{N-1}\widehat{\delta}_{{\scriptsize\textrm{sp}},0}\right\}.
\end{align}

The set $\widehat{B}_{\scriptsize\textrm{sp,imp}}$ is a basis, because its vectors are the $\textrm{GFT}$ of the vectors of the basis $B_{\scriptsize\textrm{sp,imp}}$.

\begin{definition}[Spectrum spectral impulsive representation]\label{def:spectrumgraphspfreqvecrep} The spectrum spectral impulsive representation of $\widehat{s}=\textrm{GFT}\ s$ is the coordinate vector $q_{\scriptsize\textrm{sp,imp}}$ with respect to basis $\widehat{B}_{\scriptsize\textrm{sp,imp}}$.
\begin{tcolorbox}[ams align,height=3.2cm,valign=center,title={Spectrum spectral impulsive representation},fonttitle=\bfseries\small]
\label{eqn:shatspimpulsiverep-a1}
    \widehat{s}{}&=q_0\widehat{\delta}_{{\scriptsize\textrm{sp}},0}+q_1\widehat{\delta}_{{\scriptsize\textrm{sp}},1}+\cdots+q_{N-1}\widehat{\delta}_{{\scriptsize\textrm{sp}},{N-1}}\\
    \label{eqn:shatspimpulsiverep-a2}
    &=\underbrace{\left[\begin{array}{cccc}
    \widehat{\delta}_{{\scriptsize\textrm{sp}},0}&\widehat{\delta}_{{\scriptsize\textrm{sp}},1}&\cdots&\widehat{\delta}_{{\scriptsize\textrm{sp}},{N-1}}
    \end{array}\right]}_{\widehat{D}_{\scriptsize\textrm{sp,imp}}}\underbrace{\left[\begin{array}{c}
    q_0\\
    \cdots\\
    q_{N-1}
    \end{array}\right]}_{q_{\scriptsize\textrm{sp,imp}}}
\end{tcolorbox}
\end{definition}

The spectrum spectral impulsive representation for~$\widehat{s}$ has the same coordinate vector $q_{\scriptsize\textrm{sp,imp}}$ as the spectral impulsive representation for~$s$ (see~\eqref{eqn:shatspimpulsiverep-a2} and~\eqref{eqn:sgraphfreqvectorrep-2}). It is the basis that is different. Now, it is the spectral impulse $\widehat{\delta}_{{\scriptsize\textrm{sp}},0}$ and its powers that are the basis vectors for this representation of~$\widehat{s}$.

\begin{result}[$p_{\textrm{imp}}$ and $q_{\textrm{sp,imp}}$]
\begin{align}\label{eqn:pimpandqspimp}
q_{\textrm{sp,imp}}{}&\!=\!\left(\mathcal{V}^*\right)^{-1}\!\!\textrm{GFT}^{-1}\mathcal{V} \, p_{\textrm{imp}}\textrm{  and  } p_{\textrm{imp}}\!=\!\mathcal{V} ^{-1}\!\textrm{GFT}\,\mathcal{V}^*q_{\textrm{sp,imp}}.
\end{align}
\end{result}
\begin{proof}
It follows from~\eqref{eqn:shatverteximpulsiverep-3} and~\eqref{eqn:shatspimpulsiverep-a2}, using~\eqref{eqn:graphfrequencyrepresentation-1} in result~\ref{res:DspimpandV*}.
\end{proof}

We proceed to obtain results similar to sections~\ref{subsec:polytransformfilter} and~\ref{subsec:graphztransform} for $q_{\scriptsize\textrm{sp,imp}}$. We start by associating with it a spectral polynomial transfer filter $Q_s(M)$, now in the spectral shift~$M$.
\begin{result}[$\widehat{s}$ as impulse response of $Q_s(M)$] \label{res:widehatsimppolynomialfilterQs(M)-1}
Let assumption~\ref{ass:eigunique} hold. Then $\widehat{s}$ is the impulse response of LSI filter $Q_s(M)$
\begin{align}\label{eqn:Q(M)companionshift-1}
    \widehat{s}{}&=Q_s(M)\widehat{\delta}_{{\scriptsize\textrm{sp}},0}\\
\label{eqn:shatasimprespQ(M)-1}
    Q_s(M){}&=q_0I+q_1M+\cdots+q_{N-1}M^{N-1},
    \end{align}
iff the vector of coefficients $q_{\scriptsize\textup{coef}}$ of $Q_s(M)$ is the spectral impulsive representation $q_{\scriptsize\textup{sp,imp}}$ in~\eqref{eqn:shatspimpulsiverep-a2}:
\begin{align}\label{eqn:sasimprespQ(M)-3}
q_{\scriptsize\textup{coef}}{}=\left[\begin{array}{cccc}
q_0&
q_1&
\cdots&
q_{N-1}
\end{array}\right]^T=q_{\scriptsize\textup{sp,imp}}.
\end{align}
\end{result}
Filter $Q_s(M)$ is the \textit{spectral} polynomial transform filter.

Note that equation~\eqref{eqn:sgraphfreqvectorrep-2} can be rewritten as
\begin{align}\label{eqn:shatasimprespQ(M)-1b}
   s{}&\!=\!Q_s\left(\Lambda^*\right)\!\delta_{{\scriptsize\textrm{sp}},0}\!=\! \left(\!q_0I\!+\!q_1\Lambda^*\!+\!\cdots\!+\!q_{N-1}\left(\Lambda^*\right)^{N-1}\!\right) \! \frac{1}{\sqrt{N}}1.\hspace{3mm}
\end{align}
In other words, \eqref{eqn:shatasimprespQ(M)-1b} interprets the original signal~$s$ as the impulse response of the diagonal filter $Q_s\left(\Lambda^*\right)$. From this, the next result follows.
\begin{result}[Graph signal~$\widehat{s}$ and $Q_s(M)$]\label{res:QsMandhats}
The LSI polynomial transform filter $Q_s(M)$ is alternatively given by
\begin{align}\label{eqn:QsMandhats-0}
Q_s(M){}&=\textrm{GFT}\,\textrm{diag}\left[\sqrt{N}s\right]\textrm{GFT}^{-1}
\end{align}
\end{result}
Results~\ref{res:widehatsimppolynomialfilterQs(M)-1} and~\ref{res:QsMandhats} parallel results~\ref{res:simppolynomialfilter-1} and~\ref{res:PsAands}.

\subsection{Spectral graph $z$-transform}\label{subsec:spgraphztransform}
Like $P_s(A)$ in~\eqref{eqn:sasimprespP(A)-1} led us to the $\textrm{G$z$T}$, we associate with $Q_s(M)$ a spectral graph $z$-transform ($\widehat{\textrm{G$z$T}}_{\scriptsize\textrm{sp}}$).
\begin{definition}[Spectral graph $z$-transform ($\widehat{\textrm{G$z$T}}_{\textrm{sp}}$)]\label{def:spgraphztransform} Define 
\begin{align}\label{eqn:spgraphztransform}
    \widehat{\textrm{G$z$T}}_{\textrm{sp}}{}&=\widehat{D}_{\scriptsize\textup{sp,imp}}^{-1}= \left[\begin{array}{cccc}
    \widehat{\delta}_{{\scriptsize\textrm{sp}},0}&\widehat{\delta}_{{\scriptsize\textrm{sp}},1}&\cdots&\widehat{\delta}_{{\scriptsize\textrm{sp}},{N-1}}
    \end{array}\right]^{-1}.
\end{align}
\end{definition}

\begin{result}[Fourier pairs $\widehat{\textrm{G$z$T}}^{-1}_{\textrm{sp}}$ and $\mathcal{V}^*$]\label{res:GFTandspGzT} 
\begin{align}\label{eqn:widehatGPTandVandermonde-1}
    \frac{1}{\sqrt{N}}\mathcal{V}^* \xrightarrow{\textrm{GFT}} \widehat{\textrm{G$z$T}}^{-1}_{\textrm{sp}}=\textrm{GFT}\,\frac{1}{\sqrt{N}}\mathcal{V}^*.
\end{align}
\end{result}
\begin{proof}
From~\eqref{eqn:sgraphfreqvectorrep-2}, $D_{\textrm{sp,imp}}=\frac{1}{\sqrt{N}}\mathcal{V}^*$, so, by definition~\ref{def:spgraphztransform}
\begin{align}\label{eqn:proofGPTandVandermonde-1}
\widehat{\textrm{G$z$T}}^{-1}_{\textrm{sp}}=\widehat{D}_{\textrm{sp,imp}}{}&=\textrm{GFT}\ D_{\textrm{sp,imp}}= \textrm{GFT}\frac{1}{\sqrt{N}}\mathcal{V}^*.
\end{align}
\end{proof}

$\widehat{\textrm{G$z$T}}_{\textrm{sp}}$ maps~$\widehat{s}$ into spectral $z$-transformed signals $q_{\scriptsize\textup{sp,imp}}$:
\begin{align}\label{eqn:qspimpisspGzTofhats-1}
\widehat{s}\xrightarrow{\hspace{3mm}\textup{$\widehat{\textrm{G$z$T}}_{\textrm{sp}}$}\phantom{^{-1}}} q_{\scriptsize\textup{sp,imp}}=\textup{$\widehat{\textrm{G$z$T}}_{\textrm{sp}}$}\, \widehat{s} \xrightarrow{\hspace{3mm}\textrm{$\widehat{\textrm{G$z$T}}_{\textrm{sp}}$}^{-1}} \widehat{s}
\end{align}

%
%

The $\widehat{\textrm{G$z$T}}_{\textrm{sp}}$ of~$\widehat{s}$ is the polynomial coefficient vector $q_{\scriptsize\textup{sp,imp}}$ in~\eqref{eqn:shatspimpulsiverep-a2} that defines $Q_s(M)$ in~\eqref{eqn:shatasimprespQ(M)-1} in result~\ref{res:widehatsimppolynomialfilterQs(M)-1}. To simplify notation, we use also a symbolic polynomial representation $\widehat{s}(x)$ with coefficients given by $q_{\scriptsize\textup{sp,imp}}=\widehat{\textrm{G$z$T}}_{\textrm{sp}}\,s$
\begin{align}\label{eqn:spGzTaspolys(x)}
\widehat{\textrm{G$z$T}}_{\textrm{sp}}\,\widehat{s}\approx \widehat{s}(x){}&=\left[\begin{array}{ccccc}1&x&x^2&\cdots&\end{array} x^{N-1}\right]q_{\scriptsize\textup{sp,imp}}\\
\label{eqn:spGzTaspolys(x)-2}
{}&=q_0+q_1\,x+q_2\,x^2+\cdots+q_{N-1}\,x^{N-1}.
\end{align}
The polynomial $\widehat{s}(x)$ expresses the $\widehat{\textrm{G$z$T}}_{\textrm{sp}}$ of~$\widehat{s}$ in terms of the monomial basis $B_{\scriptsize\textrm{monomial}}=\left\{1, x, \cdots, x^{N-1}\right\}$.

\textit{Spectral companion model: Canonical companion matrix and graph}. Since~$A$ and~$M$ are co-spectral (share the same spectrum), their characteristic polynomials $\Delta_A(\lambda)$ and $\Delta_M(\lambda)$ are equal. This means that we can associate with the spectral impulse representation the same companion matrix $C_{\scriptsize\textrm{comp}}$ and the same companion graph $G_{\scriptsize\textrm{comp}}$ as in sections~\ref{subsec:canonialcshift}, equation~\eqref{eqn:companionshift-9}, and~\ref{subsec:companiongraph}, respectively.


\begin{remark}[DSP representation]\label{rmk:dsprepresentations} In DSP, $\delta_0$ and its delayed replicas are impulses (in the vertex domain) and  $\widehat{\delta}_{{\scriptsize\textrm{sp}},0}$ and its delayed replicas are impulses (in the frequency domain). So
\begin{align}\label{eqn:Dspimp=identity}
D_{\scriptsize\textrm{imp}}{}&=I \textrm{   and   }
\widehat{D}_{\scriptsize\textrm{sp,imp}}=I,
\end{align}
from which (see~\eqref{eqn:impulsiverep-a2}) $s=p_{\scriptsize\textrm{imp}}$ and (see~\eqref{eqn:shatspimpulsiverep-a2}) $\widehat{s}=q_{\scriptsize\textrm{sp,imp}}$.

Further, for DSP, representation~\eqref{eqn:sgraphfreqvectorrep-2} is the Fourier representation of~$s$, i.e., decomposes the time signal~$s$ in its harmonics (the powers $\lambda^{*n}$ are the eigenvectors of $A_c$), while representation~\eqref{eqn:shatverteximpulsiverep-3} decomposes~$\widehat{s}$ in the powers of frequency vectors $\lambda^n$ (that, in DSP, are the conjugate of the harmonics).
\end{remark}
\section{Signal Representation Domains}\label{sec:signalrepdomains}
\begin{figure}[htb!]
\centering	\includegraphics[width=.35\textwidth, keepaspectratio]{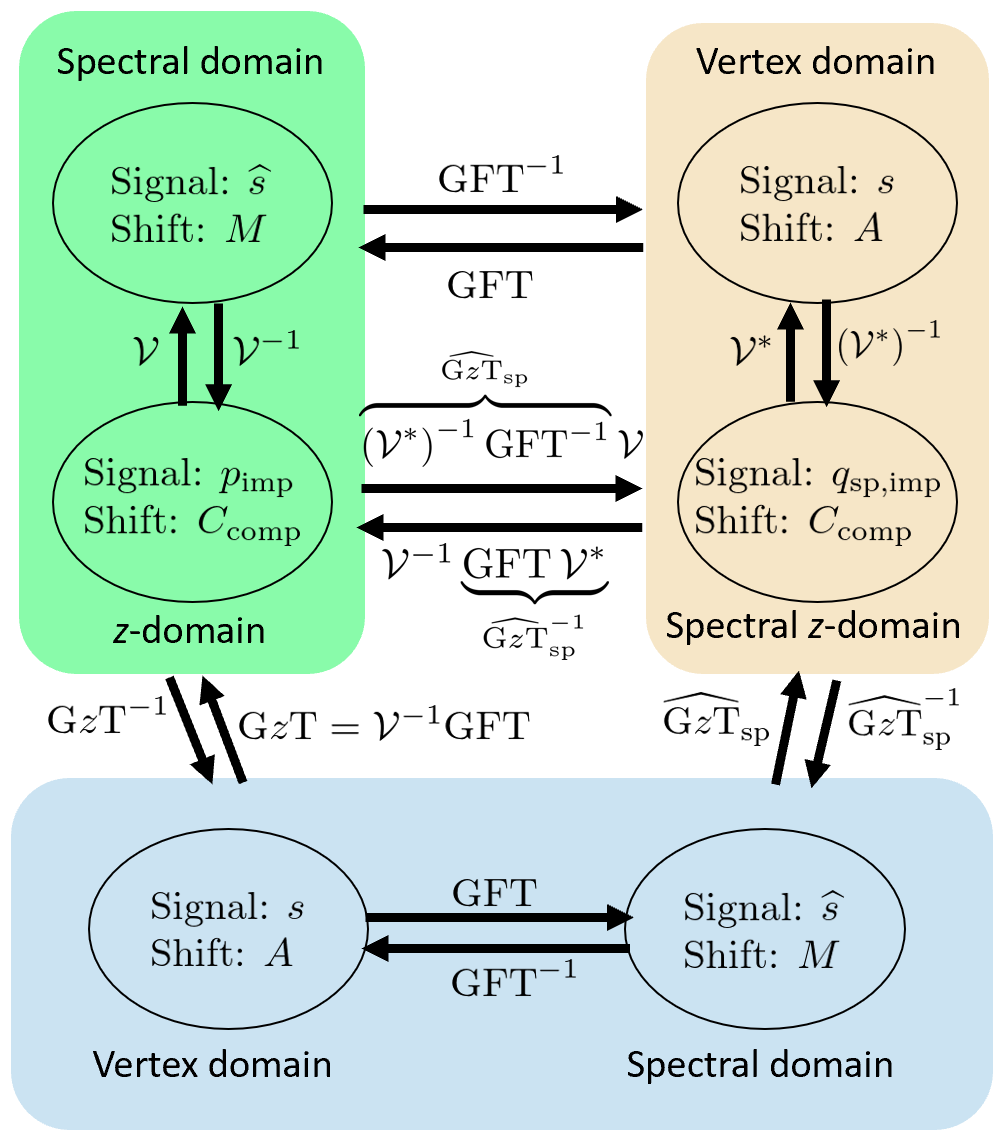}
	\caption{Graph signal domains and the transformations between them. For each domain, both the signal and shift are given.}
		\label{fig:fullpicture}
\end{figure}

In sections~\ref{sec:vertexgraphsignalrep} through~\ref{sec:repwidehats}, we discussed several signal representations. We summarize these in figure~\ref{fig:fullpicture} that illustrates the
corresponding signal domains and the transforms relating them, summarizing the main results from these sections. At the bottom, we have the standard Euclidean vertex domain signals~$s$ with its shift~$A$ and the spectral domain signals~$\widehat{s}$ with its shift~$M$. The relation between these two domains is the $\textrm{GFT}$ and its inverse $\textrm{GFT}^{-1}$. At the intermediate level, we have the two $z$-transform domains corresponding to the two impulsive representations, the vertex impulsive $z$-transformed signals $p_{\textrm{imp}}$ and the spectral impulsive $z$-transformed signals $q_{\textrm{sp,imp}}$. The graph $z$-transform $\textrm{G$z$T}$ obtains $p_{\textrm{imp}}$ from~$s$, while $\widehat{\textrm{G$z$T}}_{\textrm{sp}}$ obtains $q_{\textrm{sp,imp}}$ from~$\widehat{s}$. These two $z$-transformed signal domains reflect a number of interesting and surprising facts. Their shift is the same, the companion matrix $C_{\textrm{comp}}$, to which we associate a ``companion graph'' $G_{\textrm{comp}}$, see figure~\ref{fig:Sgraph}. In these domains, the graph eigenvalues $\left\{\lambda_n\right\}_{0\leq n\leq N-1}$ contain all needed information, since the eigenvectors derive from the graph frequency vector~$\lambda$ and its powers. At the top, we indicate that the Vandermonde matrix $\mathcal{V}$ and its conjugate relate the $z$-transformed signal domains back to the spectral and vertex domains.
%
%
%

\label{sec:dspsigreps}

\begin{figure}[htb!]
\begin{center}
	\includegraphics[width=6.5cm, keepaspectratio]{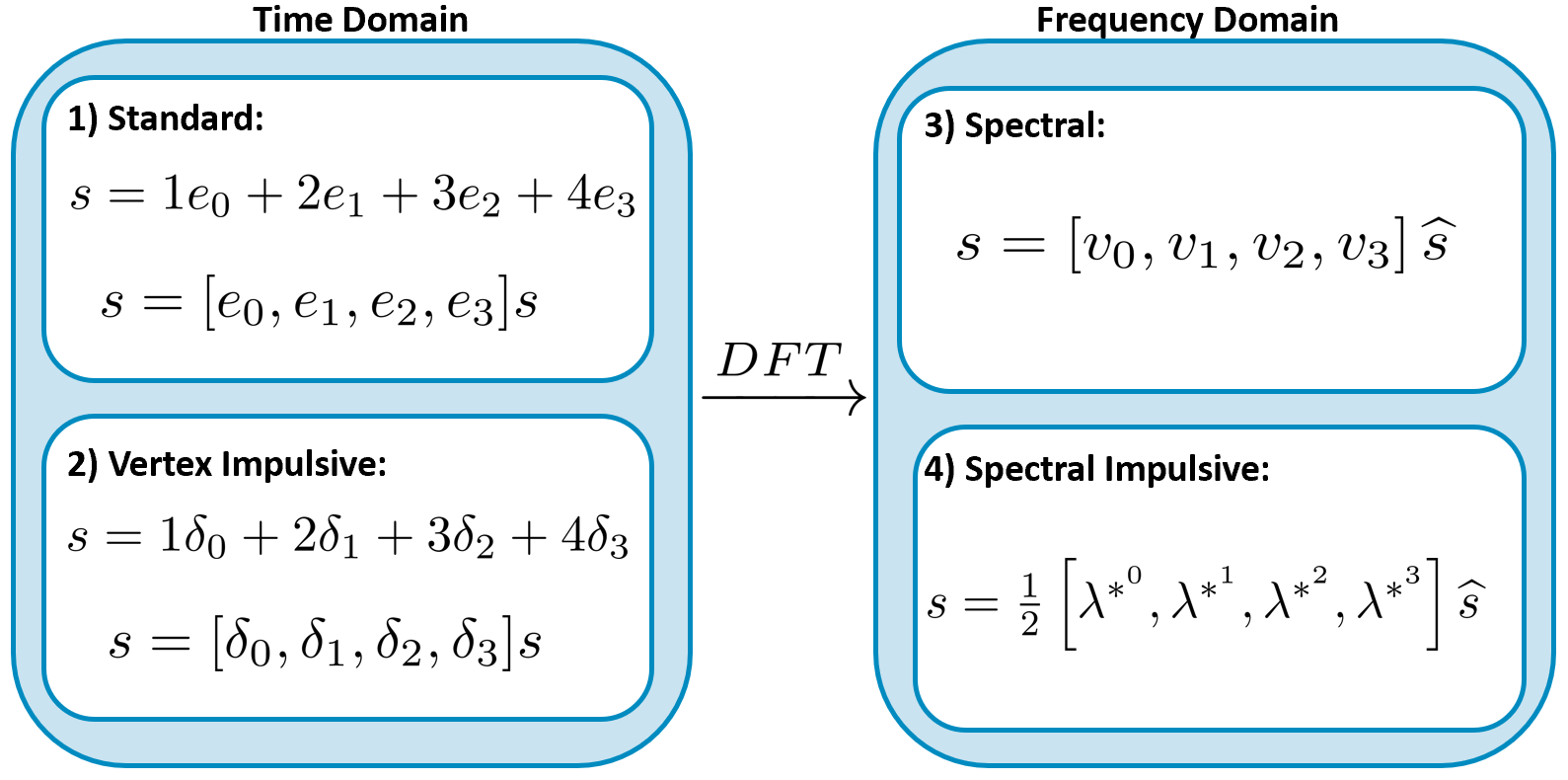}
	\caption{DSP Signal Representations:  $s = [1,2,3,4]^T$, $\widehat{s}=[5,-1+j,-1,-1-j]^T$. The standard representation and impulsive representations coincide. The Fourier and spectral impulsive representations also coincide.}
		\label{fig:dspsreps}
\end{center}
\end{figure}

With DSP the picture is much simpler. Although not usually presented this way \cite{oppenheimwillsky-1983,siebert-1986}, by reinterpreting the above GSP representations in DSP, we cast four common DSP signal representations as vector coordinatizations of the signal~$s\in\mathbb{C}^N$ with respect to choices of basis~$B$ in $\mathbb{C}^N$.
\begin{enumerate}[ labelwidth=!, labelindent=0pt]
    \item Standard: $B_E = \{e_0, e_1, \hdots, e_{N-1}\}$ is the standard or Euclidean basis and the signal representation is the vector of signal samples
    $s = [e_0, e_1, \hdots, e_{N-1}] s_{B_1} = I_N s_{B_1}=s_{B_1}$.
    \item Impulsive: $B_{\textrm{imp}} = \{\delta_0, \delta_1, \hdots, \delta_{N-1}\}$ is the basis of the impulse and its delayed replicas. Since in this case, $D_{\textrm{imp}}=I_N$, the signal representation is
    $s = [\delta_0, \delta_1, \cdots, \delta_{N-1}] p_{\textrm{imp}} = I_N p_{\textrm{imp}}$, and $p_{\textrm{imp}}=s$.
    \item Spectral: $B_{\textrm{Fourier}} = \{v_0, v_1, \hdots, v_{N-1}\}$ is the basis of the eigenmodes or harmonics and the signal representation is the Fourier transform of the signal
    $s = [v_0, v_1, \hdots, v_{N-1}] \widehat{s} = DFT^{H} \widehat{s}$.
    \item Spectral impulsive: \!\!$B_{\textrm{sp,imp}} \!\!=\!\! \frac{1}{\sqrt{N}}\left\{\lambda^{*^0}, \lambda^{*^1},\cdots,
        \lambda^{*^{N-1}}\right\}$. In this case, $D_{\textrm{sp,imp}}=\textrm{DFT}^H$, and the signal representation is
    $s = \frac{1}{\sqrt{N}}\left[\lambda^{*^0},\right.$ $\left. \lambda^{*^1}, \cdots, \lambda^{*^{N-1}}\right] q_{\textrm{sp,imp}} = DFT^{H} q_{\textrm{sp,imp}}$, and $q_{\textrm{sp,imp}}=\widehat{s}$.
\end{enumerate}
In DSP, the above four signal representations reduce to two distinct ones, see figure~\ref{fig:dspsreps} that illustrates this for a $N=4$ signal.
Since  $\delta_n = e_n, n = 0,\cdots, N-1$, the standard and impulsive bases and corresponding signal representations coincide, $B_E = B_{\textrm{imp}}$ and $s_{E} = p_{\textrm{imp}} = s$. Similarly, since $v_n = \frac{1}{\sqrt{N}} \lambda^{*^n}$, the Fourier and spectral impulsive bases and corresponding signal representations coincide, $B_{\textrm{Fourier}} = B_{\textrm{sp,imp}}$ and $\widehat{s} = q_{\textrm{sp,imp}} = \text{DFT}\:s$.

\begin{figure}[htb!]
\centering	\includegraphics[width=.35\textwidth, keepaspectratio]{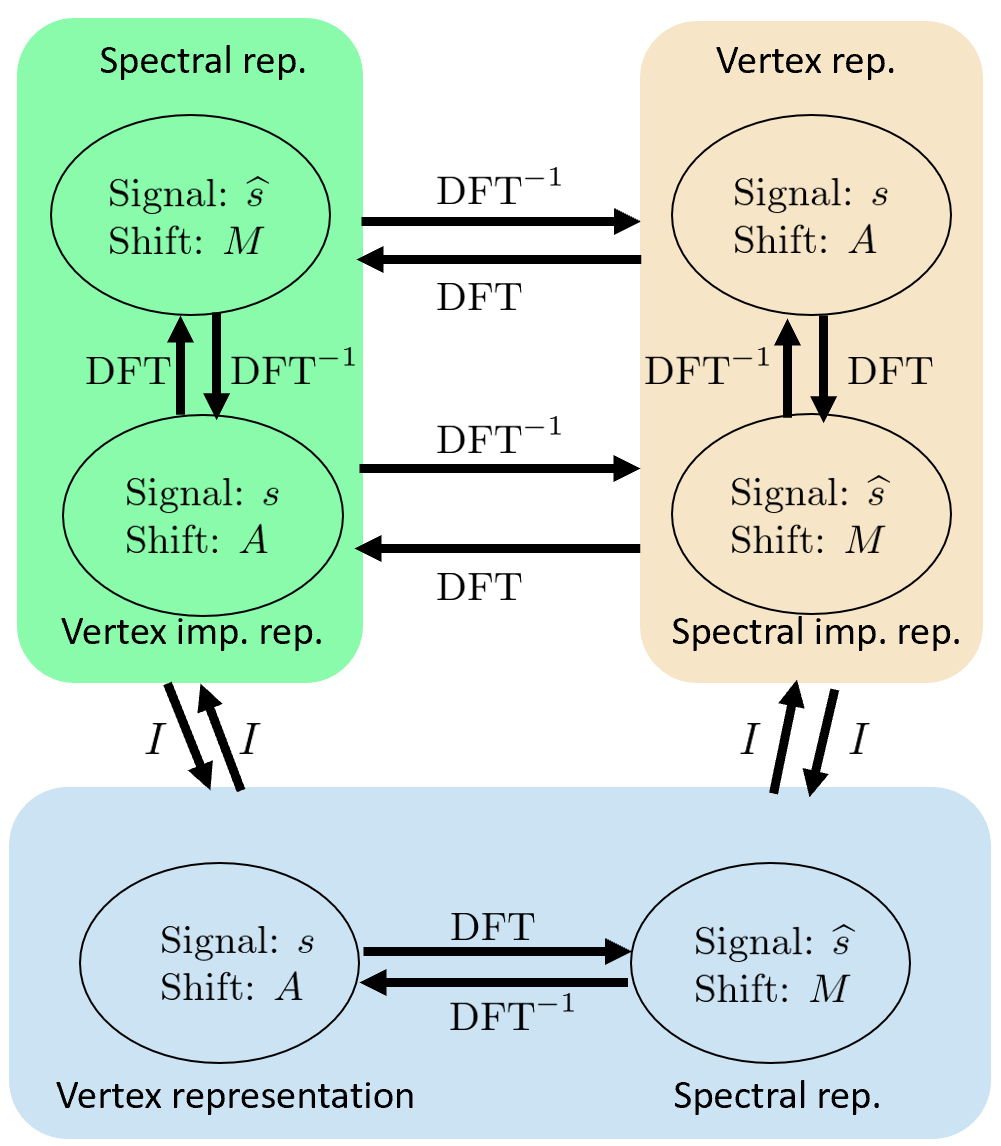}
	\caption{Figure \ref{fig:fullpicture} for DSP. The three colored regions all have the same signals and shift. In DSP, $A=M$. The three colored regions are all identical to each other in DSP.}
		\label{fig:fullpicturedsp}
\end{figure}

So, in DSP, the standard and impulsive representations can be used interchangeably as the time domain signal, $s$, and, similarly, the eigenvalue and spectral representations can be used interchangeably as the frequency domain signal, $\widehat{s}$. Also, in DSP, $A=M$\cite{shimoura-asilomar2019,shi2019graph}. Having this in mind, figure~\ref{fig:fullpicture} is much simpler with DSP as illustrated in figure~\ref{fig:fullpicturedsp}.

%
%
\section{Fast Graph Convolution With the \textrm{FFT}} \label{sec:graphconvolution}
\begin{figure}[htb!]
	\includegraphics[width=8.5cm, keepaspectratio]{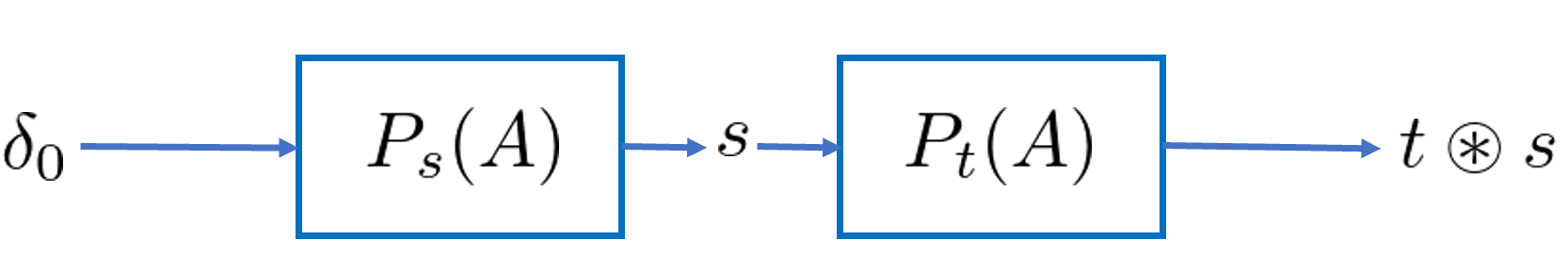}
	\caption{Convolution of graphs signals~$s$ and~$t$.}
		\label{fig:convolutionsandt}
\end{figure}

Filtering in the \textit{vertex} domain is defined in \cite{Sandryhaila:13} as the product of \textit{matrix} graph filter~$F$ with graph \textit{vector} signal~$s$. If the filter is linear shift invariant, it is a polynomial filter $P(A)$. We now consider convolution of two graph signals.
\begin{definition}[Convolution of vertex domain graph signals]\label{def:convolutiongraphsignals}
The (vertex domain) convolution of graph signals~$s$ and~$t$ is
\begin{align}\label{eqn:convtime}
t\oast s=P_t(A)\cdot P_s(A)\delta_0,
\end{align}
where $P_s(A)$ and $P_t(A)$ are the LSI polynomial transform filters for~$s$ and~$t$.
\end{definition}
Definition~\ref{def:convolutiongraphsignals} and equation~\eqref{eqn:convtime} define convolution of graph signals~$s$ and~$t$ as the impulse response of the serial concatenation of the polynomial transform filters $P_s(A)$ of~$s$ and $P_t(A)$ of~$t$. Figure~\ref{fig:convolutionsandt} illustrates this convolution.
Since polynomial filters commute, the convolution in~\eqref{eqn:convtime} commutes. The next result provides alternative ways of computing the convolution.
\begin{result}[Vertex convolution]
\label{res:convolutionsandt}
Consider graph signals~$s$ and~$t$ and their polynomial transform filters $P_s(A)$ and $P_t(A)$.  Then
\begin{align}\label{eqn:convolutionPsAtimest}
    t\oast s{}&=P_t(A)\cdot s=P_s(A)\cdot t\\
    \label{eqn:convolutionthroughspfilter-1}
   s\oast t {}&\xleftarrow{\mathcal{F}^{-1}} \sqrt{N}\,\widehat{t}\odot \widehat{s}
\end{align}
\end{result}
\begin{proof} Equation~\eqref{eqn:convolutionPsAtimest} follows from result~\ref{res:simppolynomialfilter-1} and equation~\eqref{eqn:companionshift-1}.

Equation~\eqref{eqn:convolutionthroughspfilter-1} follows by taking the GFT of both sides of~\eqref{eqn:convtime} and using the diagonalization of the transform filters in~\eqref{eqn:PsAands-0}.
\end{proof}

Equation~\eqref{eqn:convolutionPsAtimest} interprets convolution of~$s$ and~$t$ as filtering the graph signal~$s$ by a filter whose impulse response is~$t$. In the \textit{spectral} domain, equation~\eqref{eqn:convolutionthroughspfilter-1} shows that convolution of the two signals is in the spectral domain the pointwise multiplication of the GFTs of the signals. This replicates the graph Fourier filtering theorem (see equation~(27) in \cite{Sandryhaila:13}).
\subsection{Convolution of Graph Signals with the FFT}\label{subsec:directconvvertexdomain}
Equation~\eqref{eqn:convolutionthroughspfilter-1} shows that, as in DSP, we can compute convolution by finding the two GFTs~$\widehat{s}$ and~$\widehat{t}$ of the two signals, then pointwise multiplying these, and finally taking the inverse GFT of the pointwise product. Even though this replicates the DSP result, GFTs and inverse GFTs are matrix vector products that are order $N^2$, not fast operations. We show that in ``companion space,'' i.e., working with the impulsive representations, the $\textrm{G$z$T}$ of graph signals or their polynomial representation, and the polynomial transform filters of the signals, graph vertex convolution can be obtained by FFT.

\textit{Fast convolution}. Consider the $z$-transform representations  $s(x)$, $t(x)$, and $u(x)$ of~$s$, $t$, and $u=s\oast t$.
\begin{result}[GSP convolution and linear convolution]\label{res:ztransformconv}
We have
\begin{align}\label{eqn:convolutionztransf}
    u(x){}&=(s(x)\cdot t(x)) \hspace{-.3cm}\mod\! \Delta_A(x)
\end{align}
where $\Delta_A(x)$ is the characteristic polynomial of~$A$, and $u=s\oast t$ is the vector of coefficients of $u(x)$.
\end{result}
\begin{proof} The product of polynomials (in~$x$ or in~$A$) is the polynomial whose coefficients are the \textit{linear} convolution of the sequences of coefficients of the polynomials. Powers larger than $N-1$ are reduced by Cayley-Hamilton  achieved by \hspace{-.3cm}$\mod\! \Delta_A(x)$ reduction.
\end{proof}

\begin{remark}
In DSP, $A^n=A^{n\hspace{-.1cm}\mod N}$. This is the wrap-around effect, or ``time-aliasing'' in DSP. In DSP, the coefficient of the power of~$N$ is added to the coefficient of the power of $N\!\!\mod N\! =\! 0$, the coefficient of $N+1$ is added to the power of $N+1\!\!\mod N\!=\! 1$, and so on. For a generic graph in GSP, there is also ``vertex-aliasing,'' but it is not one-to-one like in DSP. The coefficient of the power of~$N$ (and higher powers) is scaled differently and added to lower powers from~$0$ to~$N-1$ as per Cayley-Hamilton.
\end{remark}
Result~\ref{res:ztransformconv} and
 equation~\eqref{eqn:convolutionztransf} are a fast convolution of the two graph signals when given their $z$-transforms. The  \textit{linear} convolution of the sequences of coefficients of $P_s(A)$ and $P_t(A)$ is computed by fast Fourier transform (FFT).  The \hspace{-.185cm}$\mod$ \hspace{-.175cm} reduction is computed by (fast) polynomial division, $O(N)$ operations  \cite{Knuth81}. Result~\ref{res:ztransformconv} is  very pleasing. It evaluates GSP LSI convolution using the FFT, an intrinsically DSP algorithm.

\begin{remark}
An interesting question is when are linear and circular (vertex) convolution equivalent in GSP. From \eqref{eqn:convolutionztransf}, we see that, in GSP, if the degree of the product polynomial $s(x)t(x)$ is
not greater than $N-1$, then
$u(x){} =(s(x)\cdot t(x)) \mod\! \Delta_A(x) = s(x)\cdot t(x)$. Linear and circular convolution are equivalent and the reduction by \!\!\!\!$\mod \Delta_A(x)$ produces no effect. This is the same condition for when linear and circular convolution are equivalent in DSP. In practice, one may want to pad with zeros either or both of $s(x)$ and $t(x)$ to get faster processing.
\end{remark}
\begin{figure*}[htb!]
\centering	
\includegraphics[width=17cm, keepaspectratio]{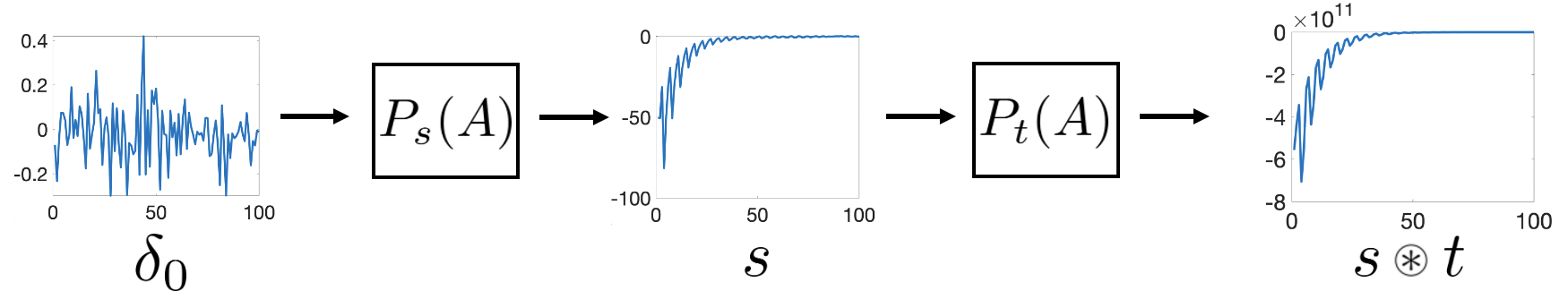}\\
\includegraphics[width=17cm, keepaspectratio]{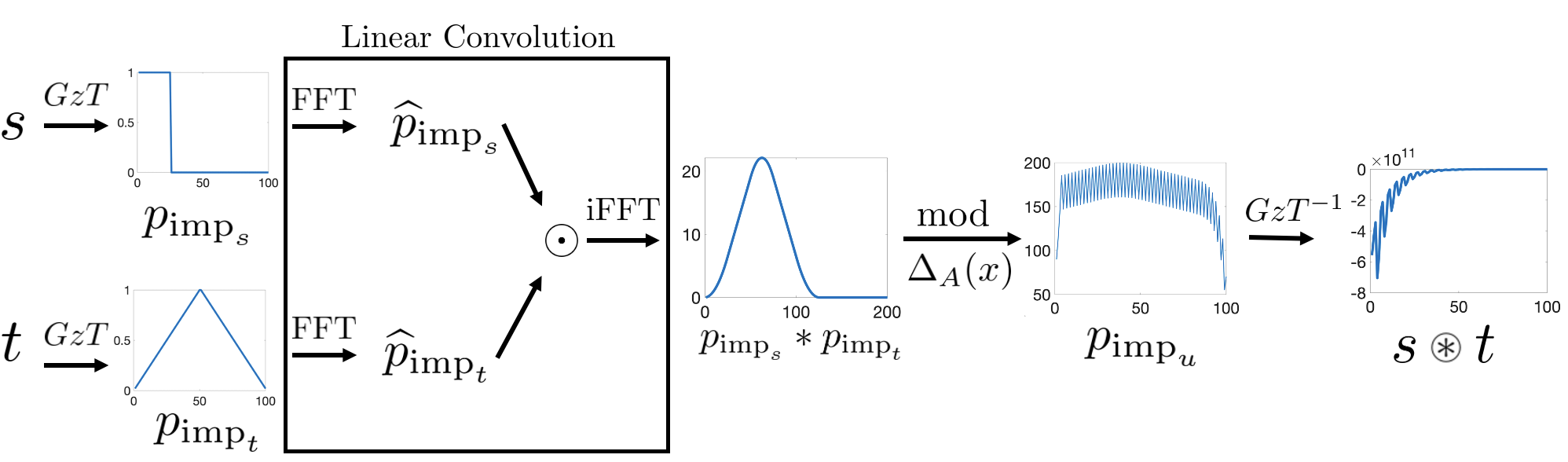}
	\caption{Example of circular convolution of $s$ and $t$ using both polynomial filtering (above) and the FFT (below) for the 100 node directed ladder graph in figure \ref{fig:ladderdir12}. Both methods produce the same $s \textcircled{$*$} t$.
	}
	\label{fig:fftexample}
\end{figure*}
An example of the convolution of graph signals using both polynomial filtering (figure~\ref{fig:convolutionsandt}) and the FFT for an expanded version (100 nodes) of the directed ladder graph in figure~\ref{fig:ladderdir12} is shown in figure~\ref{fig:fftexample}. The steps in the figure illustrate the several representations and transforms introduced in sections~\ref{sec:GSPcompanionmodel}, \ref{sec:vertexgraphsignalrep}, \ref{sec:signalrepdomains}. On top, we compute the convolution by equation~\eqref{eqn:convtime} in definition~\ref{def:convolutiongraphsignals}. From left to right, we start with the vertex impulse $\delta_0$ (obtained by $\textrm{GFT}^{-1}$ of a flat impulse in the spectral domain), going through the polynomial transform filters $P_s(A)$ and $P_t(A)$ to get $t\oast s$. At the bottom, we illustrate the fast convolution in equation~\eqref{eqn:convolutionztransf}.  We compute the $\textrm{G$z$T}$ of~$s$ and~$t$ to obtain $p_{\textrm{imp}_s}$ and $p_{\textrm{imp}_t}$. These are linearly convolved and reduced by \!\!\!\!$\mod\Delta_A(x)$ to obtain $p_{\textrm{imp}_u}$. A final $\textrm{G$z$T}^{-1}$ gets the circular convolution $s\oast t$.

The polynomial coefficient vectors for signals~$s$ and~$t$ are:
\begin{inparaenum}[1)]
\item the first 25 entries of $p_{\textrm{imp}_s}$ are~1 and the remaining 75 entries are~0; and
\item the entries of $p_{\textrm{imp}_t}$ were chosen as a triangle signal, with the first half going from $\frac{1}{50}$ to 1 with a step of $\frac{1}{50}$ and then the remaining to back down to $\frac{1}{50}$ with a step of $-\frac{1}{50}$.
\end{inparaenum}
Comparing the two plots on the right of figure~\ref{fig:fftexample}), we conclude that different methods lead to the same result for $s \oast t$, with a maximum pointwise magnitude difference (due to roundoff errors) between them of 0.15 (with convolution result values of order  $10^{11}$).
\section{Conclusion}\label{sec:conclusion}
The paper introduces the \textit{companion} graph signal model defined by a \textit{companion} shift and a \textit{companion} graph. These are canonical in the sense that every directed or undirected graph based signal model can be transformed into this canonical model by a graph $z$-transform that we define in the paper. It is obtained from impulsive representations of graph signals. The \textit{companion} graph signal model reflects many of the characteristics associated with the cyclic graph model of time signals and DSP. We show that, in the companion model, convolution of graph signals is fast convolution that is performed with the DSP FFT.

%
%

	\bibliographystyle{ieeetr}
	\bibliography{refs,sampling}
\end{document}